\documentclass[amsmath,11pt,amssymb,nofootinbib,prd]{revtex5}
\usepackage[hyperfigures,pdfstartview=FitH,bookmarksopen=false,colorlinks=false,pdfborder={0 0 0}]{hyperref}     
\usepackage{amsmath}
\usepackage{amsfonts}
\usepackage{amssymb}
\usepackage{amsthm}
\usepackage{empheq}
\usepackage{color}
\usepackage{amsmath}
\usepackage{tensor}

\def\bea{\begin{eqnarray}}
\def\eea{\end{eqnarray}}

\def\psib{\boldsymbol \psi}
\def\Phib{\boldsymbol \delta \boldsymbol \phi }

\def\bsm{\left( \!\begin{smallmatrix}}
\def\esm{\end{smallmatrix} \!\right)}

\newcommand{\smallWidthLeft}{244pt}
\newcommand{\smallWidthRight}{224pt}
\newcommand{\thirdWidthLeft}{162.667pt}
\newcommand{\thirdWidthRight}{142.667pt}

\newcommand{\be}{\begin{equation}}
\newcommand{\ee}{\end{equation}}

\def\cH{\mac{H}}

\def\d{\mathrm d}

\newcommand{\mac}[1]{\mathcal{#1}}

\begin{document}
\allowdisplaybreaks
\begin{titlepage}

\title{Cosmological Perturbations Through a Non-Singular Ghost-Condensate/Galileon Bounce}

\author{Lorenzo Battarra}
\email[]{lorenzo.battarra@aei.mpg.de}
\author{Michael Koehn}
\email[]{michael.koehn@aei.mpg.de}
\author{Jean-Luc Lehners}
\email[]{jlehners@aei.mpg.de}
\affiliation{Max Planck Institute for Gravitational Physics (Albert Einstein Institute), 14476 Potsdam, Germany}
\author{Burt A. Ovrut}
\email[]{ovrut@elcapitan.hep.upenn.edu}
\affiliation{Department of Physics, University of Pennsylvania,\\ 209 South 33rd Street, Philadelphia, PA 19104-6395, U.S.A.}

\begin{abstract}

\vspace{.3in}
\noindent We study the propagation of super-horizon cosmological perturbations in a non-singular bounce spacetime. The model we consider combines a ghost condensate with a Galileon term in order to induce a ghost-free bounce. Our calculation is performed in harmonic gauge, which ensures that the linearized equations of motion remain well-defined and non-singular throughout. We find that, despite the fact that near the bounce the speed of sound becomes imaginary, super-horizon curvature perturbations remain essentially constant across the bounce. In fact, we show that there is a time close to the bounce where curvature perturbations of all wavelengths are required to be momentarily exactly constant. We relate our calculations to those performed in other gauges, and comment on the relation to previous results in the literature.  
\end{abstract}
\maketitle

\end{titlepage}
\tableofcontents

\section{Introduction}

In the 20th century, the biggest discovery in cosmology was the realization that our universe is expanding. No other finding compelled physicists to such a revision in their thinking about the cosmos: the universe, it was then realized, is an evolving entity. The observed expansion of the universe entailed the idea that, going back in time, the universe emerged from a big bang singularity, at which time and space were supposed to begin. According to the singularity theorems of Penrose and Hawking \cite{Hawking:1969sw}, such a singularity is unavoidable in the context of general relativity, as long as certain assumptions are met. In particular, in a flat universe, a big bang singularity is unavoidable as long as {\it the null energy condition is satisfied}. This remains  true even in inflationary cosmologies: a theorem by Borde, Guth and Vilenkin \cite{Borde:2001nh} shows that inflation does not remove the big bang singularity. This is problematical since the predictions of inflation are known to be highly sensitive to the physical conditions at the onset of an inflationary stage.

The big bang singularity can also be seen as a manifestation of the breakdown of classical general relativity coupled to standard matter sources. For several decades now the hope has been entertained that a quantum theory of gravity will resolve this singularity; perhaps linking the expanding phase of the universe to a prior contracting phase \cite{Ashtekar:2006rx,Craps:2007ch,Smolkin:2012er}, perhaps joining it onto a previous fully quantum regime \cite{Hartle:2008ng}, or demonstrating the emergence of space and time from a non-geometric phase \cite{Craps:2005wd,Kleinschmidt:2009cv,Koehn:2011ph}. Unfortunately, despite many advances, there does not exist a fully understood quantum resolution of the big bang at present.

However, it is clear that such a resolution will eventually be required. A completely different approach to this problem has been presented within the context of ekpyrotic \cite{Khoury:2001wf, Donagi:2001fs, Khoury:2001bz, Khoury:2001zk} and cyclic \cite{Steinhardt:2001st,Lehners:2008vx} cosmologies. These cosmologies envisage a classical contracting universe prior to the big bang. In their initial formulation, within the context of heterotic M theory \cite{Lukas:1998yy,Lukas:1998tt,Lehners:2006pu}, the transition region between contraction and subsequent expansion was not fully elucidated at small scales due to poorly understood string physics at high energy. That is, although offering a philosophical change in point of view, there remained a potential singularity at the moment of the big bang. However, this was resolved in New Ekpyrotic Cosmology \cite{Buchbinder:2007ad,Creminelli:2007aq,Buchbinder:2007tw, Buchbinder:2007at}. These theories allow for a completely non-singular bounce within the context of classical effective field theory by introducing matter {\it violating the null energy condition}--thus evading the above singularity theorems. The original models accomplished this via a real scalar field with specific higher-derivative self-interactions, the so-called ghost condensate \cite{ArkaniHamed:2003uy}.

These considerations motivate a more general study of classically non-singular bounces, where even at the classical level the evolution through a bounce remains under control. Such classical bounces can be described in an effective field theory framework. In order to obtain a reliable non-singular bounce in a flat universe, one must suppose the existence of a matter component that can violate the null energy condition without however leading to dangerous instabilities. Examples of such matter components include scalar fields with ghost condensation \cite{ArkaniHamed:2003uy} and Galileon Lagrangians \cite{Nicolis:2008in}, and we will focus on a model of this type in the present paper. A subset of the present authors recently showed that such models can also be embedded in supergravity \cite{Koehn2013b}. This construction provided an explicit demonstration that supersymmetry and violations of the null energy condition are compatible, and provides the motivation for the present work.  

In the present paper, we are interested in the question of cosmological perturbations of such non-singular bounce spacetimes. Specifically, we would like to address the question of what happens to a set of curvature perturbations of various wavelengths going into the bounce: Will they emerge in the expanding phase, or are there catastrophic instabilities? Do the perturbations get amplified or suppressed? Does their spectrum get modified by the bounce dynamics? Since cosmological perturbations provide our main source of information about the very early universe, the answers to these questions may help to determine whether our universe underwent a bounce in our past. Indeed, we briefly discussed the evolution of cosmological perturbations through a non-singular bounce in \cite{Koehn2013b}. We showed that in commonly used gauges, such as the constant scalar field gauge \cite{Gao2011}, the equation evolving perturbations through the bounce, as well as the definition of the gauge itself, becomes undefined at the bounce when the Hubble parameter $H$ goes to zero. This seemed to preclude a well-defined solution. However, we pointed out that were one to work in a gauge--such as harmonic gauge \cite{Xue2013}--that is well-defined through the bounce, a consistent calculation of the evolution of perturbations should be possible. In this paper, we will show that this is indeed the case. Furthermore we prove, by directly studying the gauge invariant curvature perturbations ${\cal{R}}$, that the physical perturbations can be computed in any gauge--as long as appropriate care is taken with boundary conditions at the bounce.

What we find is that curvature perturbations on super-horizon scales (generated either via the entropic mechanism \cite{Notari:2002yc,Finelli:2002we,Lehners:2007ac,Buchbinder:2007ad, Buchbinder:2007tw,Li:2013hga,Qiu:2013eoa,Fertig:2013kwa,Ijjas:2014fja}, or via a preceding contracting matter phase \cite{Wands:1998yp,Finelli:2001sr}) remain essentially constant as they go through the bounce. At first sight this may appear surprising, since in the model that we are considering the speed of sound becomes imaginary in the vicinity of the bounce, signalling a gradient instability. However, the duration of the bounce is too short for long wavelength modes to be affected by it. Moreover, we can show analytically that at a time close to the bounce the curvature perturbations ${\cal R}_k$ (of all wavelengths) must be momentarily constant. That is, there is a time close to the bounce when $\dot{\cal R}_k=0.$ There is, however, some wavelength-dependence in the total amount of amplification across the bounce. Specifically, even though large-scale modes remain constant to high accuracy, as the wavelength is reduced and becomes comparable to the horizon size at the onset of the bounce, we find a small amplification. For even smaller wavelengths, our present classical description is inappropriate, as one should describe sub-horizon fluctuations using quantum field theory on our classical background spacetime. We will leave this interesting question for upcoming work.

\section{The model}
 
A non-singular bounce can only occur in a flat universe if the null energy condition (NEC) is violated. Such a violation was long thought to always be accompanied by the appearance of ghosts, i.e. kinetic (time-derivative) terms with the wrong sign. Classically, such a theory would then be unstable, as wild fluctuations would correspond to a {\it decrease} in energy. Quantum-mechanically, theories with ghosts are non-sensical, as one cannot even define a vacuum for such theories. Thus, the lesson is that ghosts must be avoided at all costs. Interestingly, it came to be realized over the last few years that certain theories with specific higher-derivative terms can allow one to violate the NEC without, however, leading to perturbative ghost instabilities. Examples of such theories are the ghost condensate \cite{ArkaniHamed:2003uy} and Galileons \cite{Nicolis:2009qm}. In this paper, we will investigate a specific bounce model that is built precisely using these theories. This model was constructed--within the context of $N=1$ supergravity--in \cite{Koehn2013b}, where the details of the construction are provided (see also \cite{Koehn:2012ar,Farakos:2012qu,Koehn:2012te,Koehn:2013hk}).
In fact, our model is closely related to a bounce model constructed by Cai, Easson and Brandenberger in \cite{Cai:2012va}. For other directly related works see \cite{Qiu:2011cy,Easson:2011zy,Qiu:2013eoa,Osipov:2013ssa}. 

Our starting Lagrangian is of the form \cite{Koehn2013b} 
\begin{eqnarray}
\mathcal{L} & = & \sqrt{-\bar{g}} \left(\frac{M_P^2}{2}\bar{R} + \bar{P}(\bar{X}, \bar\phi) + \bar{g}( \bar\phi) \bar{X}\,  \Box \bar\phi \right) \;, \\
\bar{X} & \equiv & - \frac{1}{2} \bar{g} ^{ \mu \nu} \partial _{\mu} \bar\phi\, \partial _{\nu} \bar\phi \;,
\end{eqnarray}
where $M_P$ denotes the reduced Planck mass. We are also using the definitions 
\begin{eqnarray}
\bar{P}(\bar{X}, \bar\phi) & = & \bar{K}( \bar\phi) \bar{X} + \bar{T}( \bar\phi) \bar{X} ^{2}  \;,\\
\bar{K}( \bar\phi) & = & 1 - \frac{2}{ \left(1+2 \bar\kappa \bar{\phi}^2 \right) ^2} \;,\\
\bar{T}( \bar\phi) & = & \frac{t_b}{ \left(1+2 \bar\kappa \bar{\phi}^2 \right) ^2} \;,\\
\bar{g}( \bar\phi) & = & \frac{g_b}{ \left(1+2 \bar\kappa \bar\phi^2 \right) ^2} \; .
\end{eqnarray}
Here $\bar\kappa$ (of mass dimension $-2$) parameterises the width of the bounce in scalar field space, while the dimensional constants $t_b$ (of mass dimension $-4$) and $g_b$ (of mass dimension $-3$) set the scales of the ghost condensate and Galileon terms respectively. Note that $\bar{P}(\bar{X},\bar\phi)$ in \cite{Koehn:2012ar,Koehn2013b} contains a potential energy $V(\bar\phi)$. This is important in discussing the physics prior to and following the bounce--but does not contribute significantly during the bounce itself. Hence, we can ignore it. In the present paper, we are interested in bounces where the ghost condensate dominates over the Galileon term. For this reason, it is useful to re-scale the fields according to
\begin{eqnarray}
\bar{g}_{\mu\nu} &=& t_b M_P^2 g_{\mu\nu}, \label{rescaling1}\\ \bar\phi &=& M_{P} \phi.\label{rescaling2}
\end{eqnarray}
This re-scaling has the effect of setting the scale of the ghost condensate to unity. The Lagrangian now reads
\begin{eqnarray}
\frac{1}{t_b M_P^4}  \mathcal{L} & = &  \sqrt{-g} \left(\frac{R}{2} + P(X, \phi) + g( \phi) X\,  \Box \phi \right) \;, \label{eq:Lagrescaled}
\end{eqnarray}
where $\bar{X}=X/t_b$ and the kinetic functions have simplified to
\begin{eqnarray}
P(X, \phi) & = & K( \phi) X + T( \phi) X ^{2}  \;,\\
K( \phi) & = & 1 - \frac{2}{ \left(1+2 \kappa \phi^2 \right) ^2} \;,\\
T( \phi) & = & \frac{1}{ \left(1+2 \kappa \phi^2 \right) ^2} \;,\\
g( \phi) & = & \frac{g_b}{t_b M_P}\frac{1}{ \left(1+2 \kappa \phi^2 \right) ^2} \; ,
\end{eqnarray}
with $\kappa = \bar\kappa M_P^2.$ Thus we can perform our numerical analysis using the Lagrangian \eqref{eq:Lagrescaled}, while it remains simple to transform our results to any ghost condensate scale $t_b$ via the formulae \eqref{rescaling1} - \eqref{rescaling2}.

For the background, we will assume a flat Friedmann-Lemaitre universe. In the ``physical'' time coordinate $t_{p}$, this is given by
\begin{equation}
ds ^{2} = - \d t_p ^2 + a ^2(t_p) \delta _{ij} \d x ^{i} \d x ^{j} \ .
\label{burta}
\end{equation}
However, in anticipation of our perturbative calculation we will transform to a ``harmonic'' time $t$,
which is related to the physical time by
\begin{equation}
d t_p = a(t) ^3 \, d t \;.
\end{equation}
It follows that the Friedman-Lemaitre metric becomes
\begin{equation}
ds ^{2} = - a ^{6}(t)\, \d t ^2 + a ^2(t) \delta _{ij} \d x ^{i} \d x ^{j} \ .
\label{burtb}
\end{equation}
We note that, with this coordinate  choice, the metric satisfies
\begin{equation}
\Gamma^{\mu}=g^{\rho\sigma}\Gamma^{\mu}_{\rho\sigma}=0 \ .
\label{burtc}
\end{equation}
This is a useful property when we calculate metric and scalar perturbations later in the paper--and which the metric \eqref{burta} expressed in physical time did not satisfy. For any given metric, coordinates satisfying \eqref{burtc} are called harmonic coordinates \cite{Garfinkle:2001ni,Pretorius:2004jg}. Harmonic coordinates are not unique--if $x^{\mu}$ are harmonic, then so are $y^{\mu}$=$x^{\mu}+\xi^{\mu}(x)$ as long as 
\begin{eqnarray} \label{eq:ctransf1}
 - (\xi ^{t}) ^{\prime \prime} + a ^{4} \nabla^2 \xi ^t & = & 0 \;,\\ \label{eq:ctransf2}
- \xi''  + a ^{4} \nabla^2 \xi & = & 0 \;.
\end{eqnarray}
However, the specific harmonic coordinates associated with \eqref{burtb} are ``natural'', in the sense that \eqref{burtb} easily matches ekpyrotic boundary conditions at the beginning of the bounce phase.
We emphasize that although our calculations are most transparent in harmonic time $t$--and many of our results are expressed in this variable--the physical interpretation of dynamical quantities, such as the time scale over which the bounce occurs, is most readily understood in the physical time $t_{p}$. We will make the conversion to physical time when necessary.

In the background
\begin{eqnarray}
ds ^{2} & = & - a ^{6}(t)\, \d t ^2 + a ^2(t) \delta _{ij} \d x ^{i} \d x ^{j} \;, \label{burte}\\
\phi & = & \phi(t) \; \label{burtf}
\end{eqnarray}
the equations of motion become
\begin{eqnarray}\label{bgharm1}
3 \cH ^2  &=&  - a ^{6} P + P_{,X} \phi ^{\prime 2} - 3 g\cH \frac{ \phi ^{\prime 3}}{a ^6} + \frac{1}{2} g_{,\phi} \frac{ \phi ^{\prime 4}}{ a^6} \\
\label{bgharm2}
-2 \cH' + 3 \cH ^2  &=&  a ^{6} P + g \frac{ \phi ^{\prime 2}}{a ^{6}} \left( \phi'' - 3 \cH \phi' \right) + \frac{1}{2} g_{,\phi} \frac{ \phi ^{\prime 4}}{a ^6} \\
\label{bgharm3}0 &=&  P_{,XX} \phi ^{\prime 2} \left( \phi'' - 3 \cH \phi' \right) + a ^{6} P_{,X \phi} \phi ^{\prime 2} + a ^6 P_{,X} \phi''  - a ^{12} P_{, \phi} \nonumber\\ &&
 -3 g \phi' \left[ \cH (2 \phi'' - 6 \cH \phi' ) + \cH' \phi' \right] + 2 g_{,\phi} \phi ^{\prime 2} ( \phi'' - 3 \cH \phi') + \frac{1}{2} g_{,\phi\phi} \phi ^{\prime 4}  \; ,
\end{eqnarray}
where $^\prime=\frac{d}{dt}$ and $\cH=\frac{a^{\prime}}{a}$. Again, physical quantities, such as the Hubble length just prior to and after the bounce, are best expressed in terms of ~$^{.}=\frac{d}{dt_{p}}$ and $H=\frac{\dot{a}}{a} (=\frac{{\cal{H}}}{a^{3}})$.

Note that we are only modeling the time around the bounce here. However, our analysis should be understood as providing a module that can be incorporated in a more complete cosmological model. 
We are assuming that the field starts off away from $\phi=0,$ in a contracting phase of the universe. Away from $\phi=0$ (more precisely, for $(1+2\kappa\phi^2)^2\gg1$), we have the approximate relations $K\approx 1, T\approx 0, g\approx 0,$ i.e. we have a standard kinetic term with negligible higher-derivative corrections. As the field moves toward $\phi=0,$ two things happen: its velocity increases due to blue shifting, and the higher-derivative terms start becoming important. At some point $K$ passes through zero (the dynamics is, however, non-singular at that moment due to the presence of the higher-derivative terms) and we are entering a ghost condensate phase. Moreover, the Galileon term contributes to the violation of the NEC and helps to induce a bounce. Near $\phi=0,$ we have the approximate relations $K \approx -1, T \approx 1, g\approx g_b/(t_bM_P),$ and the (smooth) bounce occurs close to this region of field space. We are mostly interested in the case where the Galileon term is comparatively small, i.e. $g_b \ll t_b M_P.$ In this case, the scale of the bounce is set by the scale of the ghost condensate: at the bounce $X \sim 1, \bar{X} \sim t_b^{-1}.$ This scale $t_b^{-1}$ can be freely adjusted, but must evidently lie above the scale of nucleosynthesis and below the Planck scale in order to provide a viable model. After the bounce, the universe starts expanding, while the scalar field velocity diminishes. 

\begin{figure}[htbp]%
\begin{minipage}{\smallWidthLeft} \flushleft
\includegraphics[width=\smallWidthRight]{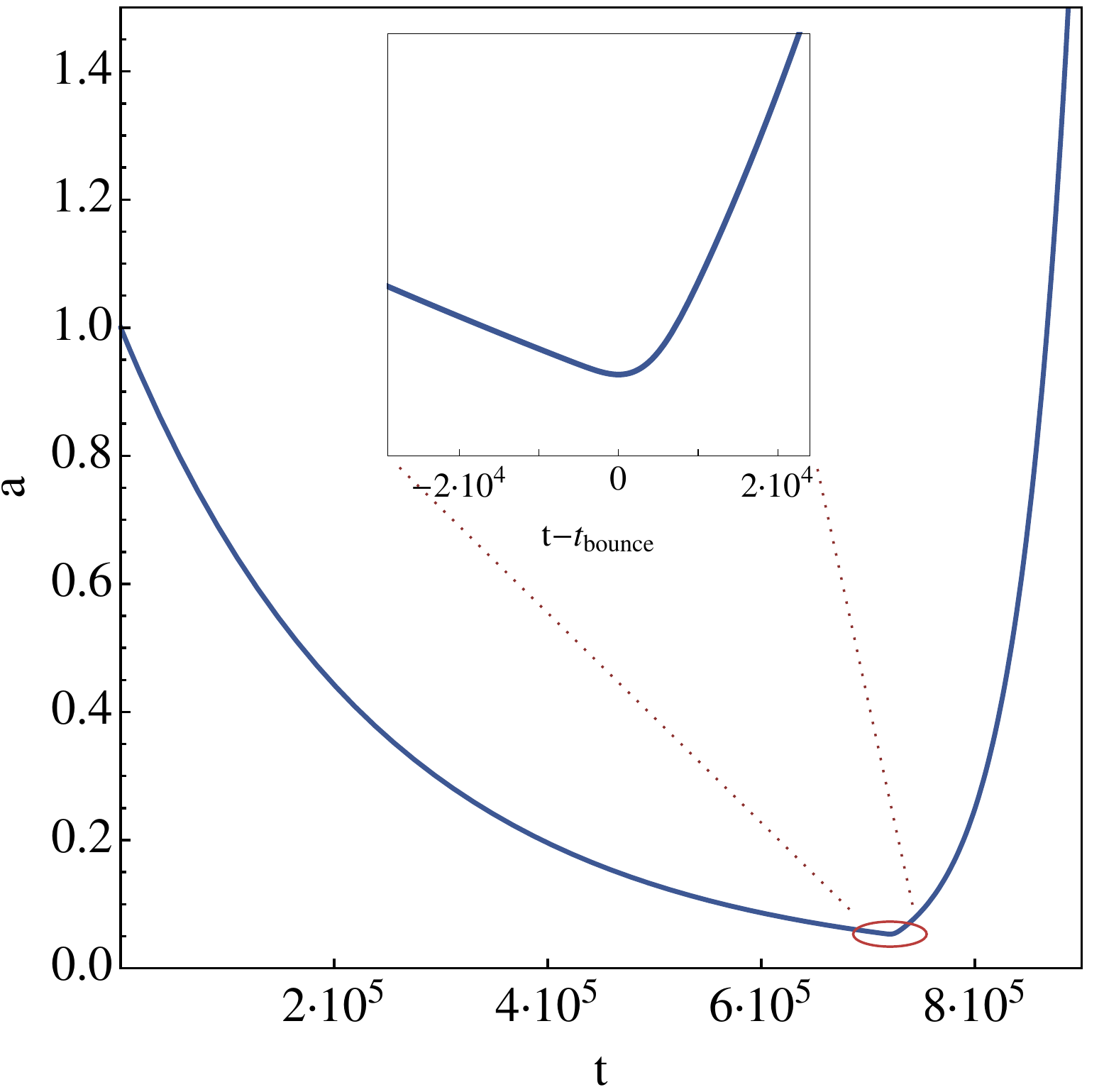}
\end{minipage}%
\begin{minipage}{\smallWidthRight} \flushleft
\includegraphics[width=\smallWidthRight]{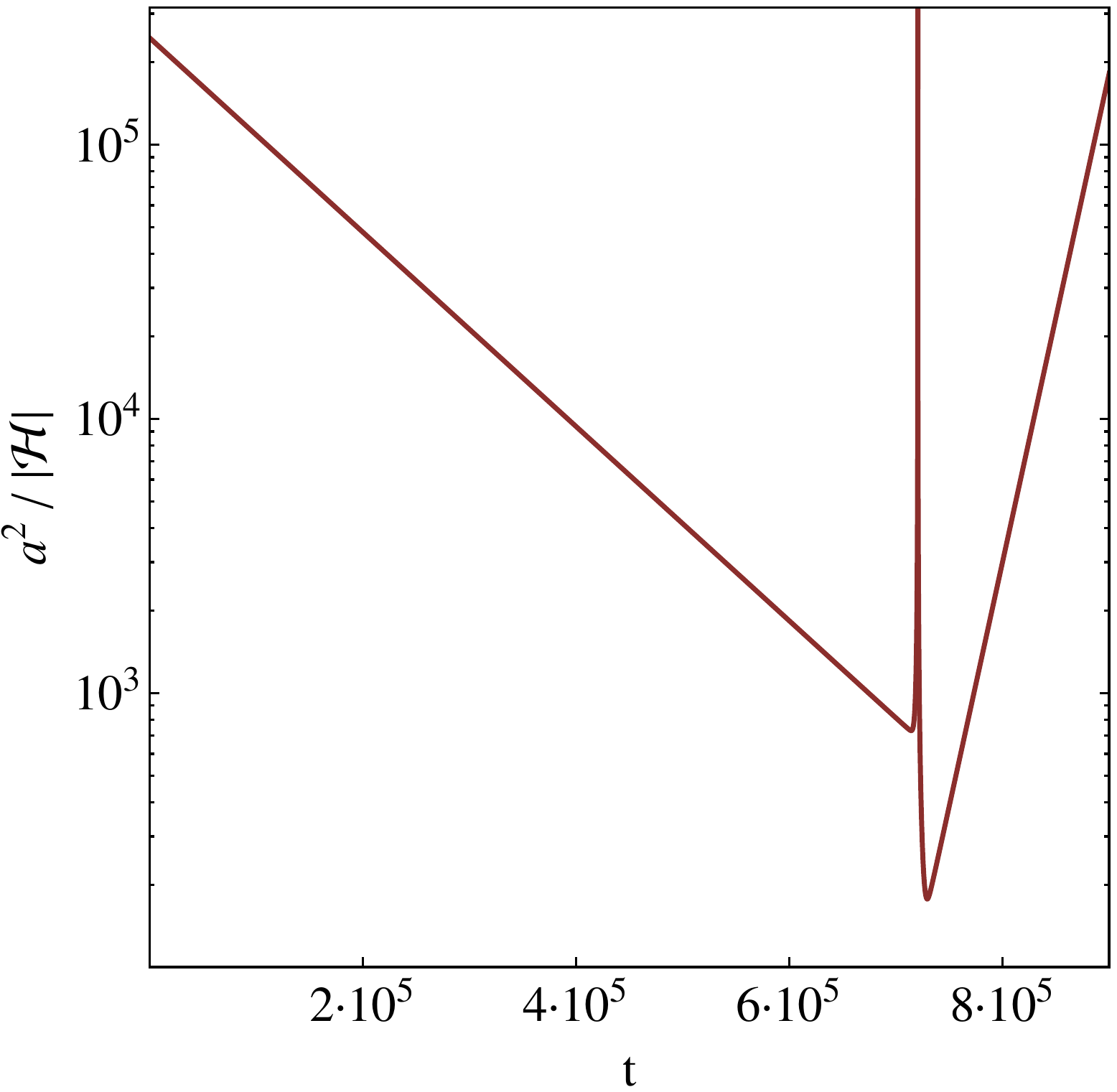}
\end{minipage}%
\caption{\label{fig:bounce1} Scale factor and comoving Hubble length for the bounce solution with parameter values $ \kappa = 1/4$, $g_b/(t_b M_P) = 1/100,$ and with initial conditions $a=1$, $ \phi = 17/2$, $ \phi^{\prime} = -10^{-5}$.}
\end{figure}

An sample solution is shown in Fig. \ref{fig:bounce1}. The plots present the evolution of the scale factor and the (absolute value of the) comoving Hubble length across the bounce, for the set of values $ \kappa = 1/4$, $g_b/(t_bM_P) = 1/100$. The initial conditions 
are $a=1$, $\phi = 17/2$, $ \phi^{\prime}= -10^{-5} (= \dot{ \phi})$. The inset shows a magnification of the scale factor around the time of the bounce, and confirms that the bounce occurs in a smooth and non-singular manner. Fig. \ref{fig:bounce7} shows the corresponding graphs of the Hubble parameter and its rate of change, as functions of physical time $t_p$. The bounce is fast in the sense that it occurs on a time-scale of ${\cal O}(1)$ as measured in units of the ghost condensate length/time scale $t_b^{1/2}M_P,$ while the horizon size ($1/H$) just before and after the bounce phase is of ${\cal O}(10)$ in the same units. If, for example, we set the ghost condensate scale two orders of magnitude below the Planck scale ($t_b = \left( 10^{-2}\right)^4$), i.e. at the grand unified scale, then this means that the horizon size at the onset of the bounce is of ${\cal O}(10^5)$ Planck lengths, while the bounce lasts about ${\cal O}(10^4)$ Planck times. Thus, for such an example the classical approximation is entirely justified. 

Note that near the bounce the comoving horizon grows and becomes infinite at the moment of the bounce, since the Hubble rate is passing through zero. Thus, all perturbation modes are inside of the cosmological horizon near the bounce. This fact raises the interesting question as to how the curvature perturbations evolve across the bounce.

\section{Linear perturbations in harmonic gauge}

\subsection{Singularity of the equation of motion for the curvature perturbation}

\begin{figure}[t!]
\begin{minipage}{\smallWidthLeft} \flushleft
\includegraphics[width=\smallWidthRight]{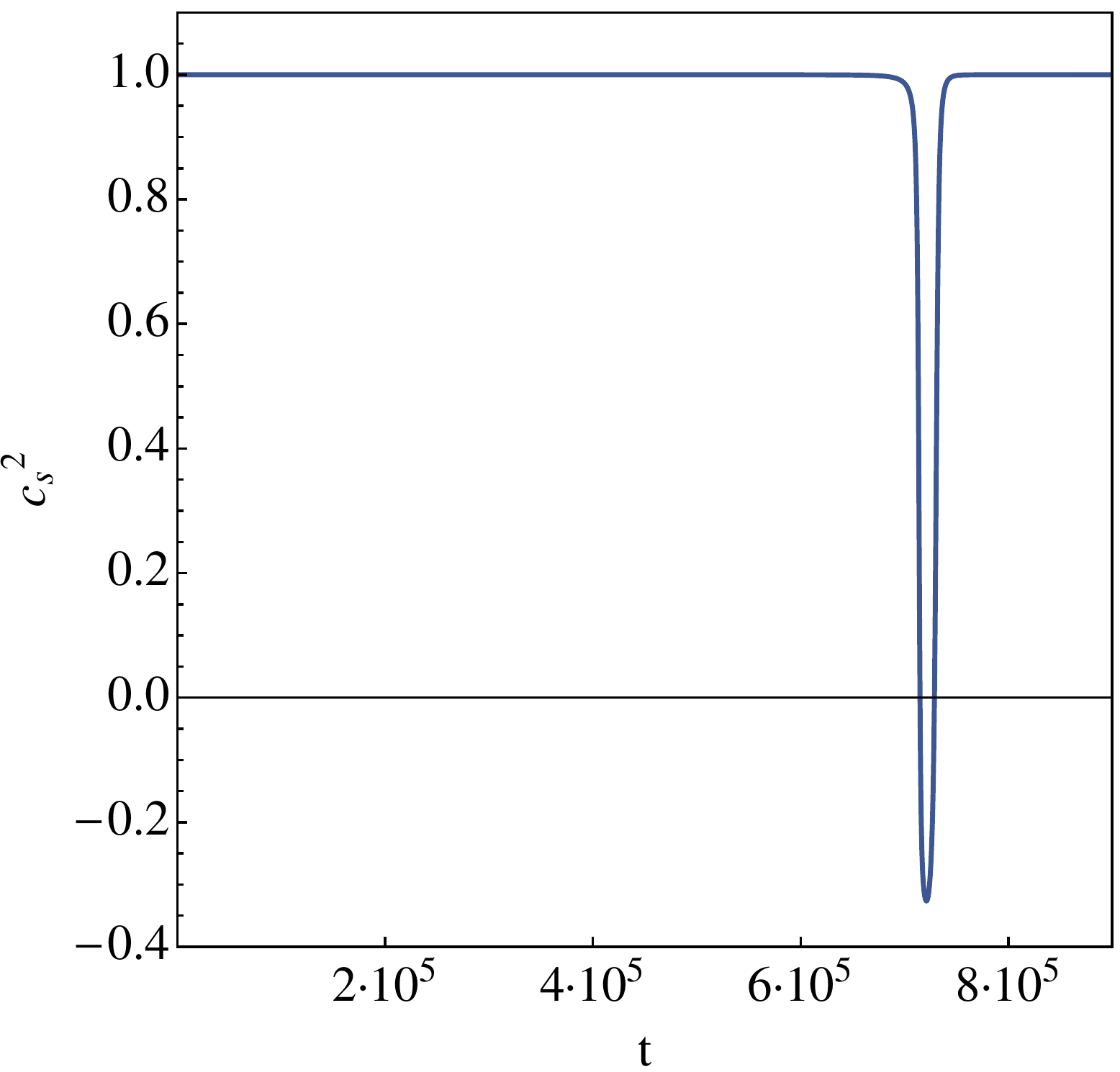}
\end{minipage}%
\begin{minipage}{\smallWidthRight} \flushleft
\includegraphics[width=\smallWidthRight]{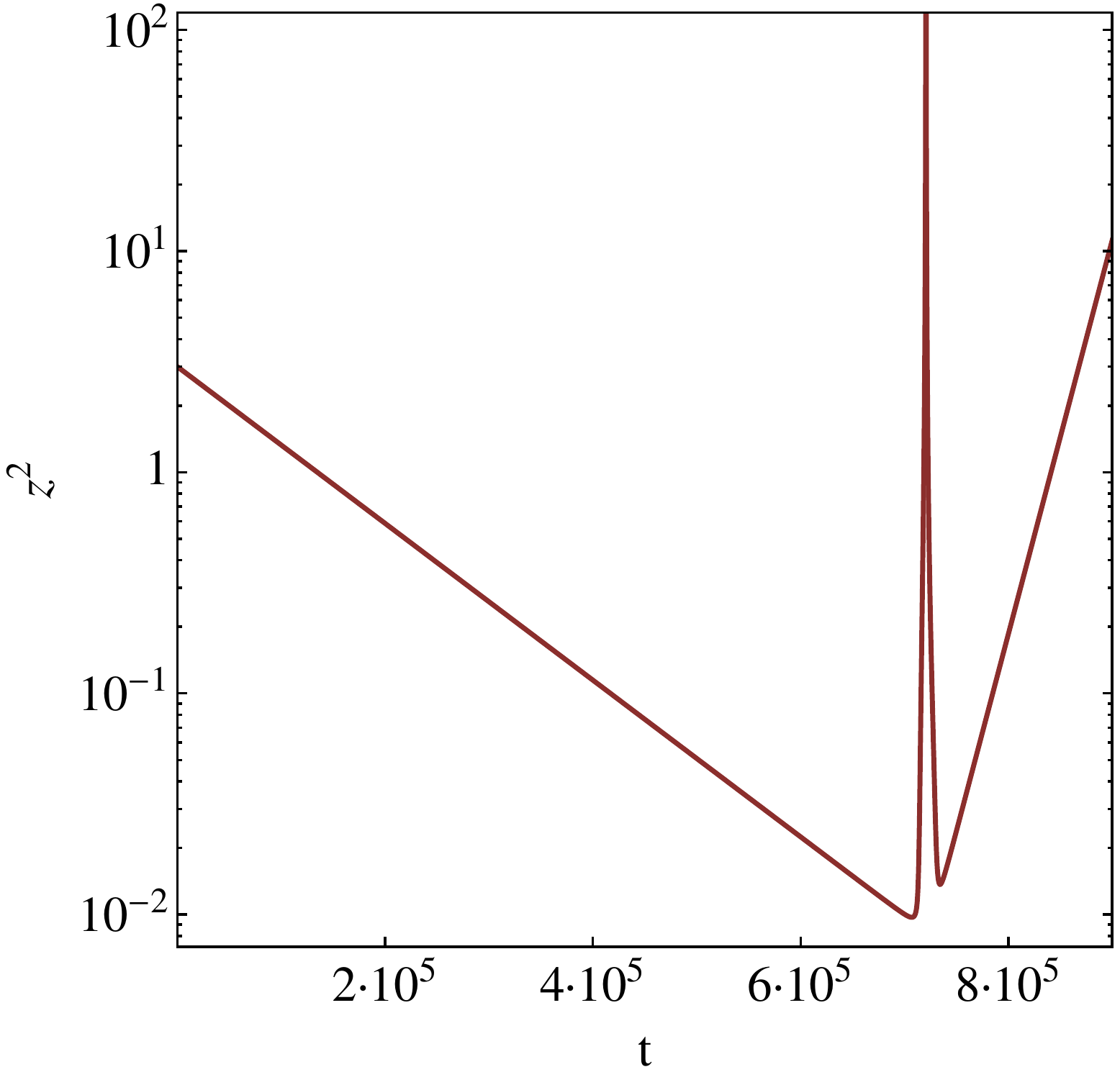}
\end{minipage}%
\caption{\label{fig:bounce3} Evolution of the speed of sound squared and of $z ^2$ in the non-singular bounce background. The positivity of $z^2$ demonstrates the absence of perturbative ghost fluctuations, while the brief period over which $c_s^2$ becomes negative indicates the presence of a gradient instability.}
\end{figure}

The (gauge--invariant) co-moving curvature perturbation $ \mathcal{R}$ satisfies the closed equation \cite{Deffayet:2010qz,Kobayashi:2010cm,Gao2011,Kobayashi:2011nu}
\begin{equation} \label{eq:conformalR}
\frac{ d ^2 \mathcal{R}}{d \tau ^2} + \frac{2}{z} \frac{ d z }{ d \tau} \frac{ d \mathcal{R}}{ d \tau} + c  _{s} ^2 k ^2 \mathcal{R} = 0 \;,
\end{equation}
where $ d \tau \equiv a ^2 d t$ is conformal time. The coefficients appearing in \eqref{eq:conformalR} are given by
\begin{eqnarray}
z ^2 & = & \frac{1}{2} \frac{ a ^2 \phi ^{\prime 2} \mathcal{P}}{ \left( \mathcal{H} + \frac{1}{2} g( \phi) \frac{ \phi ^{\prime 3}}{a ^6} \right) ^2} \;,\label{eq:z2} \\
\mathcal{P} & = &  P_{,X} - 6 g( \phi) \mathcal{H} \frac{\phi'}{ a ^6} + \frac{3}{2} g ^{2}( \phi) \frac{ \phi ^{\prime 4}}{a ^{12}} + 2 g_{,\phi} \frac{ \phi ^{\prime 2}}{a ^6} + P_{,XX} \frac{ \phi ^{\prime 2}}{ a ^{6}} \;, \label{eq:calP} \\
c_s ^2 & = &  \frac{1}{ \mathcal{P}} \left(P_{,X} + 2  g( \phi) \cH \frac{ \phi'}{ a ^6} - \frac{1}{2} g ^2( \phi) \frac{ \phi ^{\prime 4}}{a ^{12}} - 2 g( \phi) \frac{ \phi''}{a ^{6}} \right) \;.\label{eq:cs2}
\end{eqnarray}
The quantities $c_s^2$ and $z^2$ are plotted in Fig. \ref{fig:bounce3}. For completeness, we note that $z^2$ appears as the coefficient of the kinetic term of $\mathcal{R}$ in the perturbed action of our model \cite{Koehn2013b}, and thus its positivity is synonymous with an absence of ghosts. Note that $z^2$ blows up in the vicinity of the bounce, as the denominator of \eqref{eq:z2} passes through zero when $\mathcal{H} =- \frac{1}{2} g( \phi) \frac{ \phi ^{\prime 3}}{a ^6}.$ This implies that at this moment the equation for $\mathcal{R}$ becomes singular. We will discuss this-- ultimately harmless--singularity of the equation of motion for the curvature perturbation in more detail in section \ref{section:singularities}. For now, the lesson we draw from this observation is that it would be desirable to find a better, non-singular and completely reliable way to describe the evolution of perturbations across the bounce. 

Fig. \ref{fig:bounce3} also shows the evolution of the speed of sound squared $c_s^2$ of the curvature perturbations. An interesting feature is that near the bounce the speed of sound becomes imaginary, signalling a gradient instability. The presence of this instability may lead one to speculate that, during the bounce, perturbations will be strongly (perhaps catastrophically) amplified. In fact, a large amplification has been claimed in \cite{Cai:2012va} in a bounce model essentially identical to the one we are considering here (and where the singular equation for $\mathcal{R}$ was used to study the evolution of the perturbations). A counter-argument is that the bounce occurs over a very brief time period, and that it is hard to imagine how perturbations with a wavelength longer than the scale of the bounce can be much affected by the gradient instability. 

In order to resolve these issues unambiguously, we will perform our calculations in harmonic gauge where, as we will demonstrate, the evolution of the perturbations is entirely non-singular. We would like to highlight that harmonic gauge was also used recently in a paper by Xue et al. \cite{Xue2013} in the context of a simpler bounce model (which, however, contains ghosts).

\subsection{Harmonic gauge} \label{section:harmonic}

As discussed previously, our background classical fields are given in a specific harmonic gauge   by \eqref{burte}, \eqref{burtf}. Starting in that gauge, we now write the generic linearized scalar perturbations of our background fields as
\begin{eqnarray}
ds ^2 & = & - a ^{6}(1 + 2 \mathbf{A}) \d t ^2 + 2 a ^{4} \mathbf{B}_{,i}\, \d t\, \d x ^{i} + a ^2(t) \Big[ \left(1- 2 \psib \right) \delta _{ij} + 2 \mathbf{E}_{, ij} \Big] \d x ^{i} \d x ^{j} \;,\\
\phi & = & \phi(t) + \Phib(t, x) \;,
\end{eqnarray}
where, for the sake of clarity, we will denote metric and scalar field perturbations in boldface. Is this new metric still in a harmonic gauge? Varying $\Gamma^{\mu}$ with respect to these perturbations, while leaving the coordinates unchanged, we find that
\begin{eqnarray}
\delta \Gamma ^{ \mu} & = & - \frac{1}{a ^{6}} c  ^{\mu} \;,\\
c ^{t} & = & \mathbf{A}'   + 3 \psib' - \nabla ^2 \left(\mathbf{E}' - a ^{2} \mathbf{B} \right) \;, \label{eq:ct} \\
c ^{i} & = & \Big[ \left(a ^2 \mathbf{B} \right)' + a ^{4} \left( \mathbf{A} - \psib - \nabla ^2 \mathbf{E} \right) \Big] _{,i} \;. \label{eq:ci}
\end{eqnarray}
Hence, to remain in a harmonic gauge in these coordinates--that is, to satisfy \eqref{burtc}--one must set $\delta \Gamma^{\mu}=0$. Therefore, the perturbation functions must satisfy
\begin{eqnarray} \label{eq:finalPert1burt}
0 & = & \mathbf{A}' + 3 \psib' + k ^{2} \left( \mathbf{E}' - a ^{2} \mathbf{B} \right) \ , \\  \label{eq:finalPert2burt}
0&=&\left( a ^{2} \mathbf{B} \right)' + a ^{4} \left(\mathbf{A} - \psib + k ^{2} \mathbf{E} \right) \ .
\end{eqnarray}
Finally, recall that under coordinate transformations  with parameters $\xi^\mu=(\xi^t,\xi^{,i})$ satisfying the constraints \eqref{eq:ctransf1} and \eqref{eq:ctransf2}, the unperturbed metric continues to be harmonic. Under the same gauge changes, we find that the perturbations transform as
\begin{eqnarray} \label{eq:gt1}
\mathbf{A} & \rightarrow & \mathbf{A} - \xi ^{t \prime} - 3 \cH \xi ^t \;,\\ \label{eq:gt2}
\mathbf{B} & \rightarrow & \mathbf{B} + a ^{2} \xi ^t - \frac{1}{a ^2} \xi ^{\prime} \;,\\ \label{eq:gt3}
\psib & \rightarrow & \psib + \cH\, \xi ^t \;,\\ \label{eq:gt4}
\mathbf{E} & \rightarrow & \mathbf{E} - \xi \;.
\end{eqnarray}
Be that as it may, the conditions \eqref{eq:finalPert1burt}, \eqref{eq:finalPert2burt} continue to be satisfied. Thus, one can use this {\it residual harmonic gauge freedom} to fix the initial values of the perturbation functions arbitrarily, and we will do so in the next section. Solutions to the harmonic wave equations \eqref{eq:ctransf1} - \eqref{eq:ctransf2} exist for the whole bouncing background, implying that the harmonic gauge is well-defined throughout the bouncing phase. Notice also how in the variation $ \delta \Gamma ^{ \mu}$ the differential operators which act on the components of $ \xi ^{\mu}$ are non--singular at the bounce. Hence, the gauge constraints \eqref{eq:finalPert1burt}, \eqref{eq:finalPert2burt} can be imposed without problems. 

In the Appendix, we provide a detailed derivation of the linearized Einstein and scalar field equations in harmonic gauge. This analysis shows that, after a number of simplifications, one is left with the following set of perturbation equations (where the first two equations are the gauge constraints):
\begin{eqnarray} \label{eq:finalPert1}
0 & = & \mathbf{A}' + 3 \psib' + k ^{2} \left( \mathbf{E}' - a ^{2} \mathbf{B} \right)\\  \label{eq:finalPert2}
0&=&\left( a ^{2} \mathbf{B} \right)' + a ^{4} \left(\mathbf{A} - \psib + k ^{2} \mathbf{E} \right)\\
\label{eq:finalPert3}
\label{t-tBis} 0 & = & \Big( \cH'- \frac12 P_{,XX} \frac{ \phi ^{\prime 4}}{a ^{6}} + 3 g \cH \frac{ \phi ^{\prime 3}}{ a ^{6}} +\frac12\frac{g\phi'^2\phi''}{a^6} - \frac12g_{,\phi} \frac{ \phi^{\prime 4}}{a ^{6}} \Big) \mathbf{A} - k ^2\Big(a ^2 \cH +\frac{1}{2} g \frac{ \phi ^{\prime 3}}{a ^{4}} \Big) \mathbf{B}\nonumber\\
&& + 3 \left(\cH + \frac{1}{2} g \frac{ \phi ^{\prime 3}}{a ^6} \right) \psib'+ k ^2 a ^4  \psib + k ^2 \left( \cH + \frac{1}{2} g \frac{ \phi ^{\prime 3}}{a ^6} \right) \mathbf{E}' \nonumber\\
&& + \frac{1}{2} \left(P_{,X} \phi' + P_{,XX} \frac{ \phi ^{\prime 3}}{a ^{6}} - 9 g \cH \frac{ \phi ^{\prime 2}}{a ^6} + 2 g_{,\phi} \frac{ \phi ^{\prime 3}}{a ^6} \right) \Phib' \nonumber\\
&& - \frac{1}{2} \left( a ^6 P_{, \phi} - P_{, X \phi} \phi ^{\prime 2} + k ^2 g \frac{ \phi ^{\prime 2}}{a ^2} + 3 g_{,\phi} \cH \frac{ \phi^{\prime 3}}{a ^6} - \frac{1}{2} g_{,\phi\phi} \frac{ \phi ^{\prime 4}}{a ^6} \right) \Phib \;,\\
\label{eq:finalPert4}
\label{t-iBis} 0 & = & \left(\cH + \frac{1}{2} g \frac{ \phi ^{\prime 3}}{a ^6} \right) \mathbf{A} +  \psib'   - \frac{1}{2} g \frac{ \phi ^{\prime 2}}{a ^{6}} \Phib' 
- \frac{1}{2} \frac{ \phi'}{a ^6} \left( a ^6 P_{,X} - 3 g\cH \phi' + g_{,\phi} \phi ^{\prime 2} \right) \Phib \;,\\
\label{eq:finalPert5}
0 & = & \frac{1}{a ^{4}} \mathbf{E}''+k^2\mathbf E \;.
\end{eqnarray}
Note that we write the perturbation equations in Fourier space, with $k$ being the comoving wavenumber. We do not need to consider the linearized scalar field equation, as it can be derived from the equations above (as discussed in the Appendix).

\begin{figure}[t!]
\begin{minipage}{\smallWidthLeft} \flushleft
\includegraphics[width=\smallWidthRight]{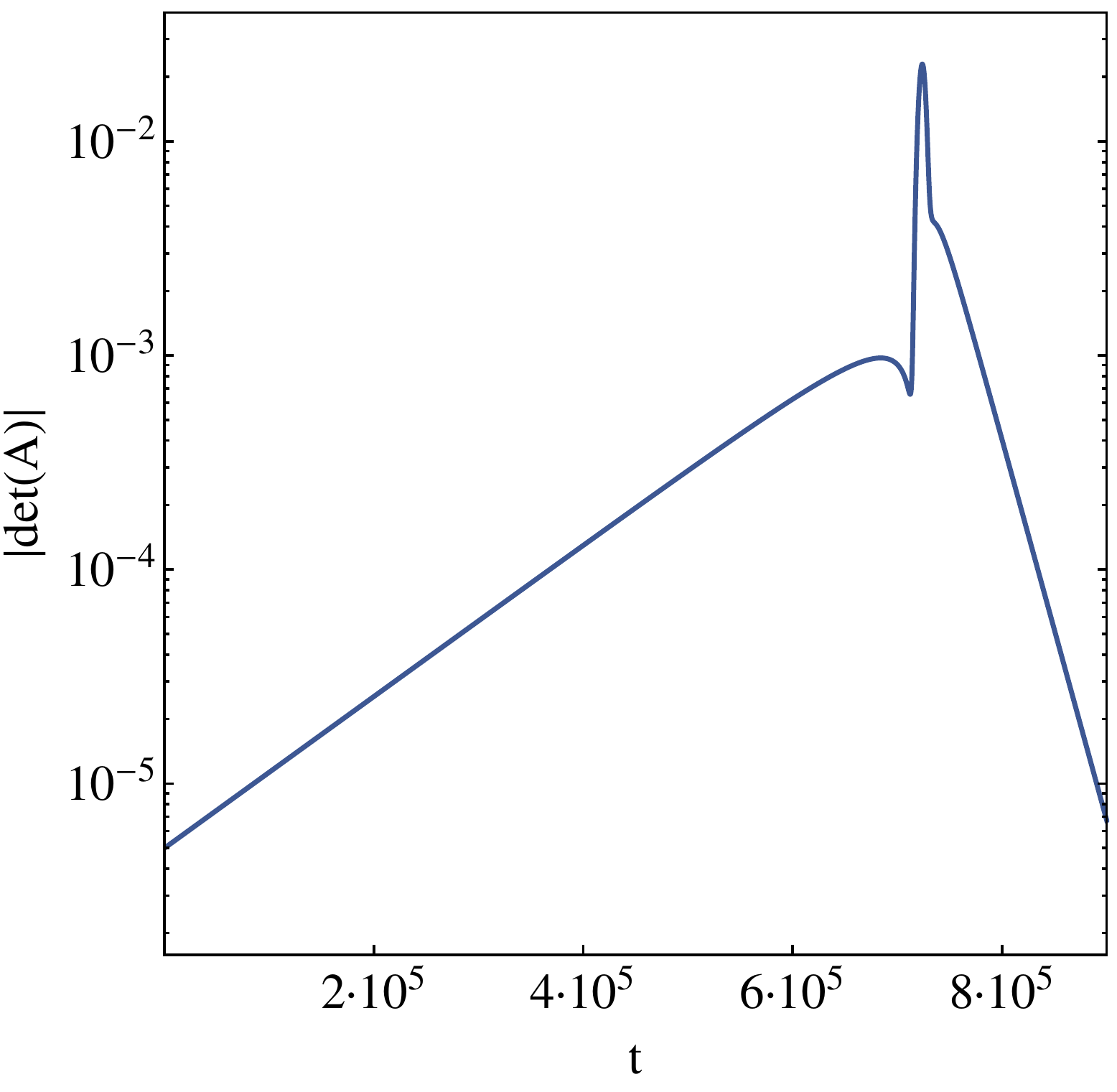}
\end{minipage}%
\begin{minipage}{\smallWidthRight} \flushleft
\includegraphics[width=\smallWidthRight]{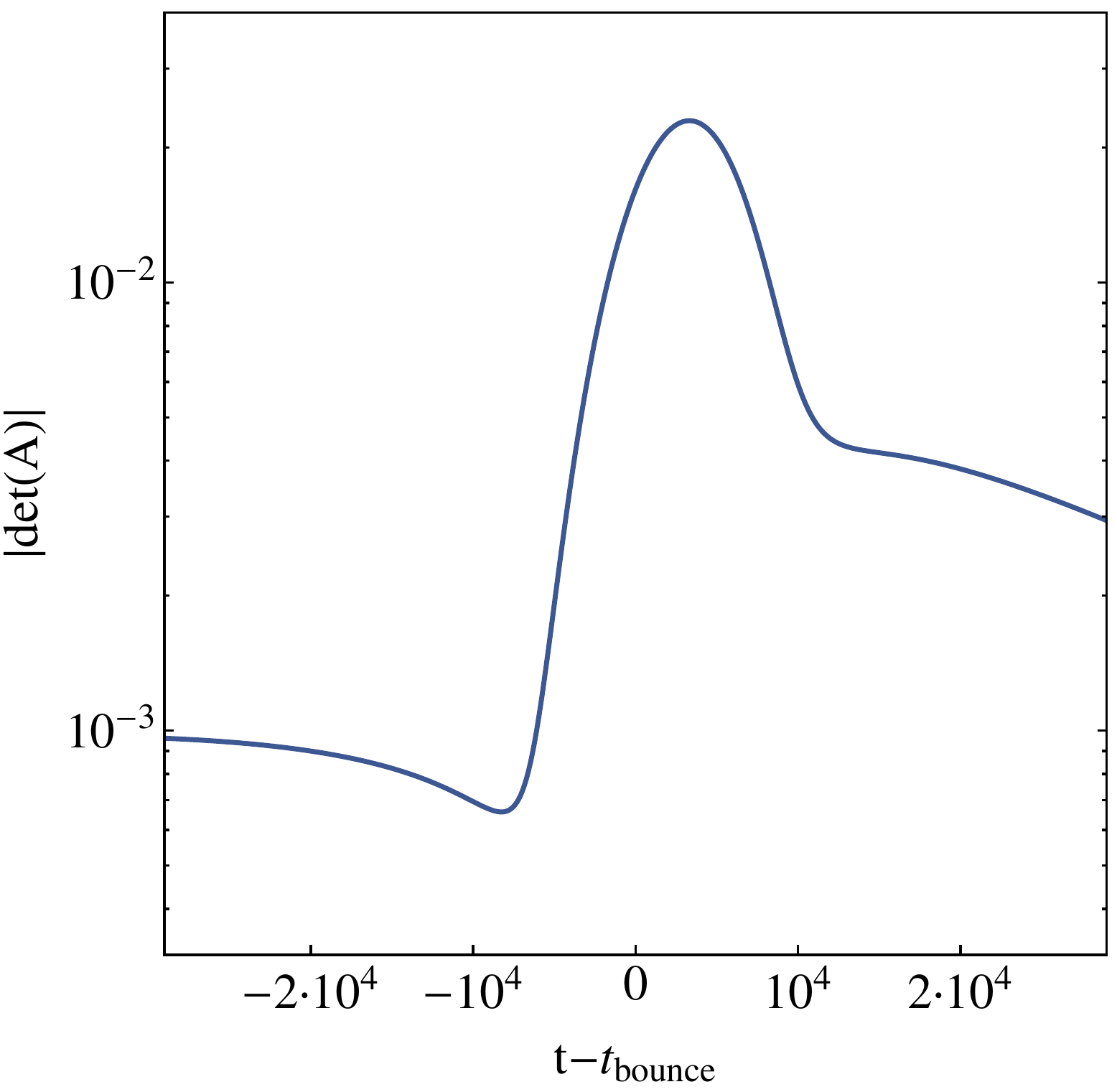}
\end{minipage}%
\caption{\label{fig:bounce2} Determinant of the matrix $A$ appearing in \eqref{eq:system}. The right panel shows a magnification around the time of the bounce. As is evident, the matrix $A$ is regular and invertible throughout, demonstrating that our system of perturbation equations is non-singular.}
\end{figure}

Notice the most crucial aspect of these equations: all coefficients are non-singular in a bounce spacetime. In particular, there are no $1/\mathcal{H}$ factors present. In fact, since this was the main motivation for using harmonic gauge in the first place, we should be more rigorous and explicitly show that the system of equations (\ref{eq:finalPert1})--(\ref{eq:finalPert5}) is entirely non--singular. This can be accomplished by re-writing it in the form
\begin{equation} \label{eq:system}
A(t) F'(t) + B(t) F(t) = 0 \;,
\end{equation}
where $F \equiv ( \mathbf{A},  \mathbf{B}, \psib,  \mathbf{E},  \mathbf{E}', \Phib) $ and $A$, $B$ are time--dependent matrices given by the background. A supplementary first order equation for $ \mathbf{E}'$ obviously has to be included in the system. A direct inspection of (\ref{eq:finalPert1})--(\ref{eq:finalPert5}) shows that the individual matrix elements in $A(t)$ and $B(t)$ are well--behaved as long as the background itself is well--behaved, which we implicitly assume. Possible divergences can then only arise if the matrix $A$ fails to be invertible. We can make sure that this does not happen by numerically evaluating the determinant of $A,$ see Fig. \ref{fig:bounce2}. The figure explicitly shows that the determinant is non-zero throughout, and this simple result implies that the evolution of the perturbation variables across the bounce is regular.


\section{Evolving through a non--singular bounce}

Having derived the linearized equations of motion in harmonic gauge, we now proceed to solve these numerically. 

\subsection{Initial conditions for the perturbations}

The main application that we have in mind in the present work are bouncing models of the universe in which nearly scale-invariant curvature perturbations are generated during the contracting phase preceding the bounce. Several mechanisms are known which can achieve this, for example the entropic mechanism (see \cite{Fertig:2013kwa,Ijjas:2014fja} for the most appealing models) or a contracting matter phase (see e.g. \cite{Cai:2011zx,Cai:2013kja}) \footnote{The entropic mechanism does not amplify gravitational waves, while a contracting matter phase does (typically with roughly the same amplitude as the scalar perturbations). Entropic models fit the Planck satellite data well \cite{Lehners:2013cka}, but would be ruled out if the claimed detection of primordial gravitational waves reported by the BICEP2 collaboration \cite{Ade:2014xna} receives independent confirmation.}. For this reason, we will choose as our initial conditions a set of classical, scale-invariant curvature perturbations. The gauge--invariant comoving curvature perturbation is defined as
\begin{equation}
\mathcal{R} \equiv \psib + \frac{\cH}{ \phi'} \Phib \;.
\end{equation}
We have the following expectation values for super-horizon modes (see \cite{Battarra:2013cha} for a detailed treatment of the quantum-to-classical transition of these perturbations)
\begin{eqnarray}
\langle \mathcal{R} ^{2} \rangle &\sim& k ^{-3} \;,\\
\langle \mathcal{R} ^{ \prime 2} \rangle &\sim& k \;.
\end{eqnarray}
Hence, our initial conditions for classical mode functions are $ \mathcal{R} \sim k ^{-3/2}$, $ \mathcal{R}' \sim k ^{1/2}$. Initial conditions for the perturbations variables $ \mathbf{A}$, $ \mathbf{B}$, $\psib$, $ \mathbf{E}$ and $ \Phib$ can be obtained by completely fixing a gauge. Indeed, the gauge conditions \eqref{eq:finalPert1}, \eqref{eq:finalPert2} are preserved by infinitesimal diffeomorphisms generated by $ \xi ^{\mu}$ satisfying the harmonic constraints (see \eqref{eq:ctransf1}, \eqref{eq:ctransf2})
\begin{eqnarray}
( \xi ^{t})'' + a ^4 k ^2 \xi ^{t} & = & 0 \;, \\
\xi'' + a ^4 k ^2 \xi & = & 0 \;.
\end{eqnarray}
The general solution of these equations can be specified by assigning arbitrary values of $ \xi ^{t}$ and $ \xi$, together with their first time derivative, at the instant $t_0$ where we specify the boundary conditions for the perturbation equations. By analyzing the gauge transformations (\ref{eq:gt1})--(\ref{eq:gt4}), it is easy to see that the values, for example, of $ \mathbf{A}$, $ \mathbf{B}$, $\psib$ and $ \mathbf{E}$ can be set to any arbitrary value at $ t = t_0$. The value of $ \Phib(t_0)$ is then obtained from the boundary condition for $ \mathcal{R}$, while $ \mathbf{E}'(t_0)$ can be obtained from $ \mathcal{R}'(t_0)$ by using the perturbation equations (\ref{eq:finalPert1})--(\ref{eq:finalPert5}). 

As an example, we can set for each Fourier mode
\begin{equation}
\mathbf{A}(t_0) =  \mathbf{B}(t_0) = \psib(t_0) =  \mathbf{E}(t_0) = 0 \;.
\end{equation}
At this reference time, where we will also take $a = 1$ to simplify the variable changes, the linearized equations (\ref{eq:finalPert1})--(\ref{eq:finalPert5}) reduce to:
\begin{eqnarray}
0 & \stackrel{t=t_0}{=} &  \mathbf{A}' + 3 \psib' + k ^{2} \mathbf{E}' \;,\\
0 & \stackrel{t=t_0}{=}  & \mathbf{B}' \;, \\ \label{eq:eqinc}
0 & \stackrel{t=t_0}{=}  & 3 \left(\cH + \frac{1}{2} g( \phi) \phi ^{\prime 3} \right) \psib'+  k ^2 \left( \cH + \frac{1}{2} g( \phi) \phi ^{\prime 3} \right) \mathbf{E}' \nonumber\\
&& + \frac{1}{2} \left(P_{,X} \phi' + P_{,XX} \phi ^{\prime 3} - 9 g( \phi) \cH \phi ^{\prime 2} + 2 g_{,\phi} \phi ^{\prime 3} \right) \Phib' \;, \\
0 & \stackrel{t=t_0}{=}  & \psib'   - \frac{1}{2} g( \phi) \phi ^{\prime 2} \Phib' 
- \frac{1}{2} \phi' \left( P_{,X} - 3 g( \phi) \cH \phi' + g_{,\phi} \phi ^{\prime 2} \right) \Phib \;, \\
0 & \stackrel{t=t_0}{=}  & \mathbf{E}'' \;.
\end{eqnarray}
So we still need to specify the values of $ \Phib$ and $ \mathbf{E}'$. Using the definition of $\mathcal{R}$ and the equations above we get the intermediate results
\begin{eqnarray}
\Phib' & \stackrel{t=t_0}{=}  & \frac{ \phi'}{ \mathcal{H} + \frac{1}{2} g( \phi) \phi ^{\prime 3}} \left\{ \mathcal{R}' - \left[ \left( \frac{ \mathcal{H}'}{ \mathcal{H}} - \frac{ \phi''}{ \phi'} \right) + \frac{1}{2} \frac{ \phi ^{\prime 2}}{ \mathcal{H}} \left( P_X - 3 g( \phi) \mathcal{H} \phi' + g_{,\phi} \phi ^{\prime 2} \right) \right] \mathcal{R} \right\} \;,\\
\psib' & \stackrel{t=t_0}{=} & \mathcal{R}' - \left( \frac{ \mathcal{H'}}{ \mathcal{H}} - \frac{ \phi''}{ \phi'} \right) \mathcal{R} - \frac{ \mathcal{H}}{ \phi'} \Phib' \; .
\end{eqnarray}
One can then obtain the value of $ \mathbf{E}' $ at $t = t_0$ from \eqref{eq:eqinc}. The initial conditions are completed by specifying
\begin{equation}
\Phib_0  =  \frac{ \phi'_0}{ \mathcal{H}_0} \mathcal{R}_0 \;.
\end{equation}

\subsection{Numerical results for the evolution of curvature perturbations} \label{section:results}

We are finally in a position to calculate the evolution of super-horizon curvature fluctuations in our non-singular bounce spacetime. Let us briefly recall the clash of the two intuitive expectations. On one hand, one would expect long-wavelength modes to be unaffected by a bounce occurring on much smaller scales. On the other hand, at the bounce all modes re-enter the horizon and a gradient instability is present, which may lead one to speculate that all modes might grow significantly across the bounce. We can expand slightly on this heuristic remark by estimating the naively expected growth due to the gradient instability. Using Eq. (\ref{eq:conformalR}), we would estimate the growth as 
\begin{equation}
\mathcal{R}_{\text{post-bounce}} \sim \textrm{exp} \left( k \int_{ c _{s} ^2 < 0} |c_s| d \tau \right) \mathcal{R}_{\text{pre-bounce}} \sim e^{ k / k_\star} \mathcal{R}_{\text{pre-bounce}}  \;.
\end{equation}
For the background considered here, numerical integration gives $k_\star \simeq 0.05$. 

\begin{figure}[t!]
\begin{minipage}{\smallWidthLeft} \flushleft
\includegraphics[width=\smallWidthRight]{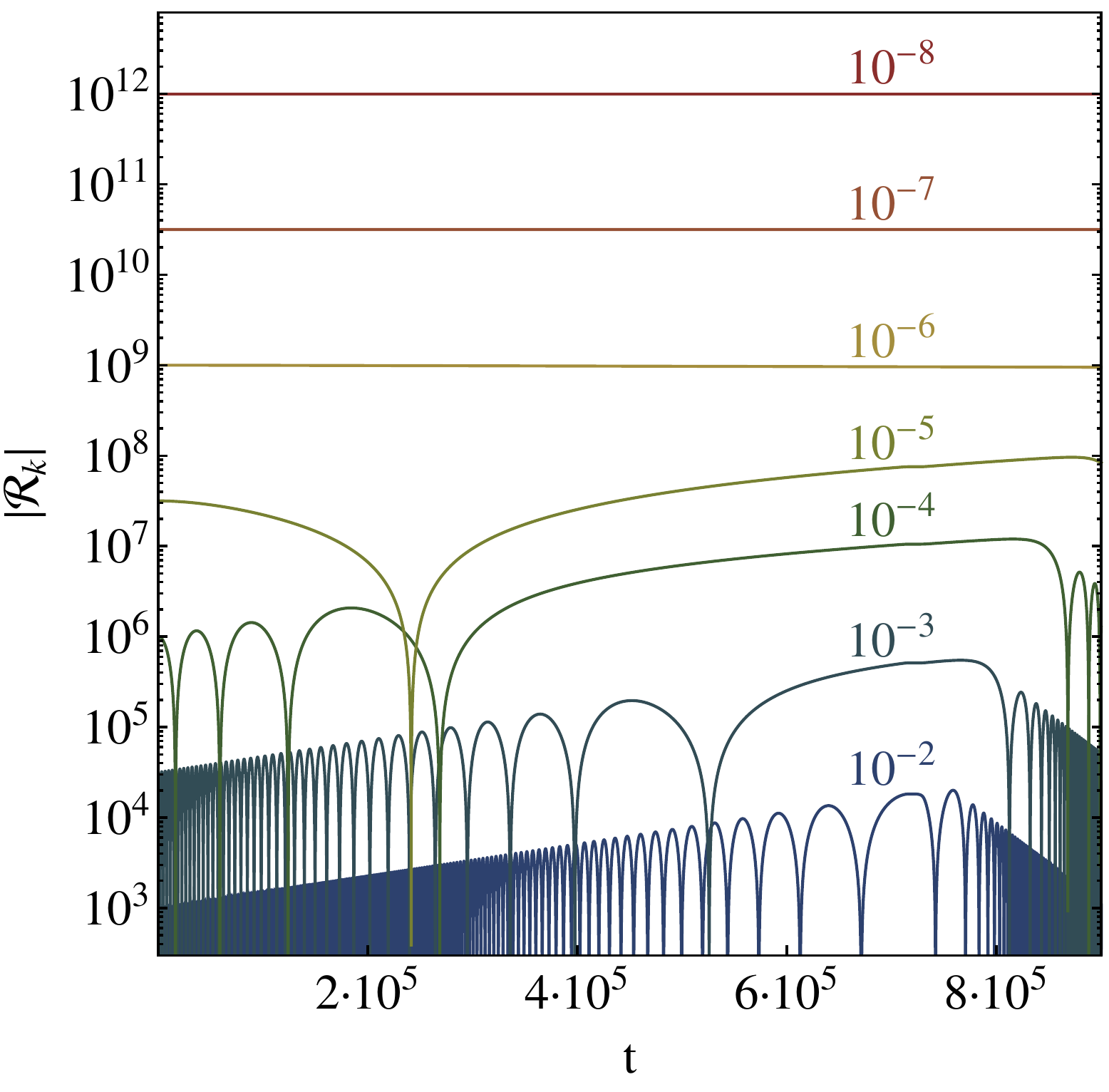}
\end{minipage}%
\begin{minipage}{\smallWidthRight} \flushleft
\includegraphics[width=\smallWidthRight]{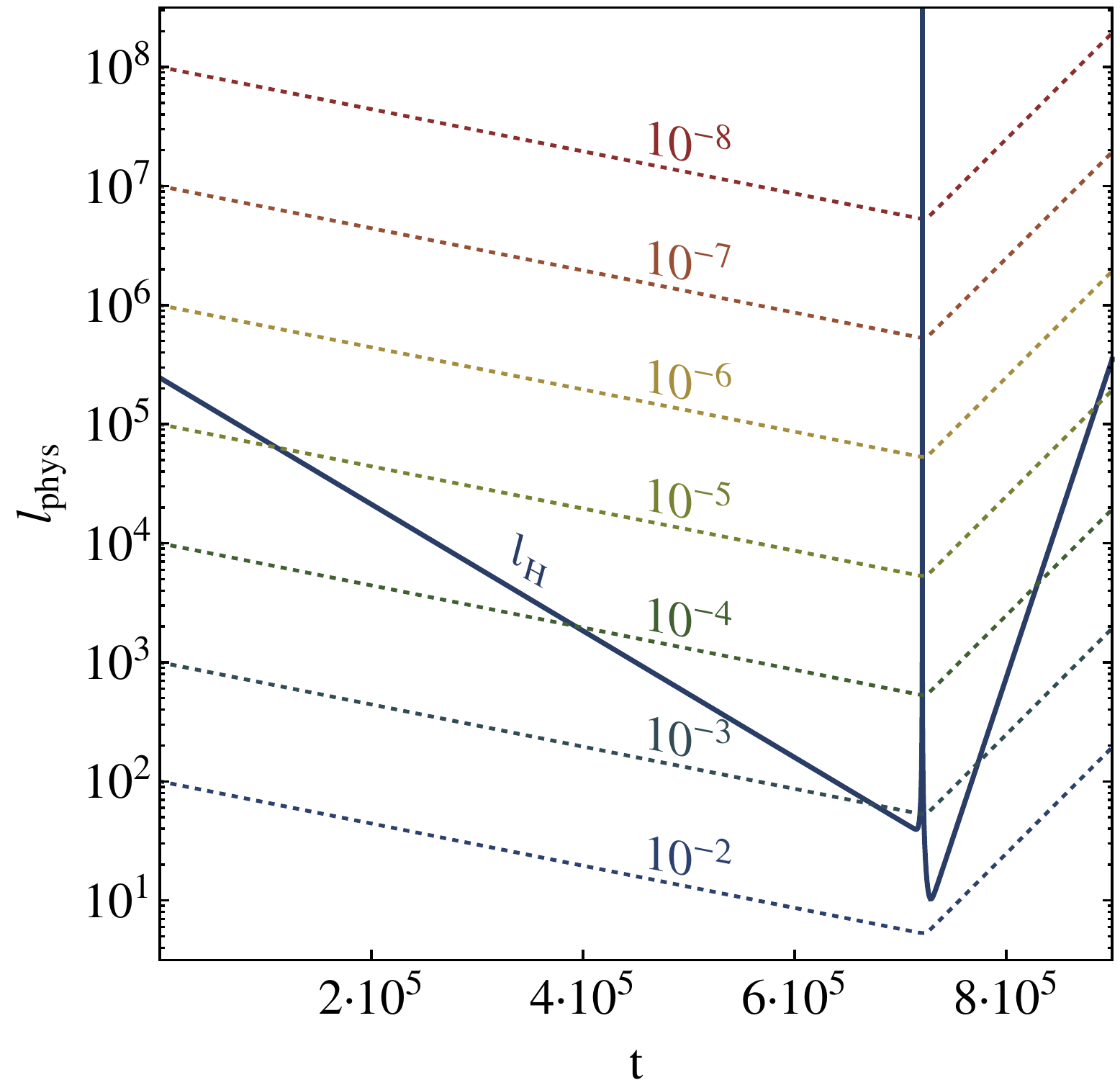}
\end{minipage}%
\caption{\label{fig:bounce5} The left-hand panel presents the evolution of the comoving curvature perturbation for various wavelengths. Long-wavelength modes are preserved essentially unchanged across the bounce, while short, initially oscillating modes, flatten out near the bounce. The right-hand panel shows both the horizon size and the physical wavelengths $a/k$ of the various perturbation modes (on a logarithmic scale) as functions of harmonic time.}
\end{figure}

The numerical evolution of the curvature perturbations is shown in Fig. \ref{fig:bounce5} for wave numbers $k$ in the range $10^{-2} - 10^{-8}.$ The long-wavelength modes $k=10^{-6},10^{-7},10^{-8}$ are everywhere super-horizon in the approach to the bounce, and their classical description is fully appropriate. As is evident from the figure, these modes remain constant with very high precision across the bounce. Thus, the intuition that the bounce is too short for such long-wavelength modes to be affected has proven correct. Despite the fact that these modes briefly enter the horizon around the time of the bounce, their amplitude remains essentially unchanged (magnification would reveal a tiny and utterly negligible enhancement of the amplitude across the bounce). This constitutes our main finding: non-singular ghost condensate/Galileon bounces preserve both the amplitude and spectrum of large-scale curvature perturbations across the bounce, and hence, if such perturbations are generated during the contracting phase, they will go through unmodified into the expanding phase of the universe.

It is, however, also of interest to study the behavior of shorter-wavelength modes. For $k=10^{-3},10^{-4},10^{-5}$ the right-hand panel of Fig. \ref{fig:bounce5} shows that they leave the horizon just before the bounce phase. Thus, around that time a classical description is fairly appropriate. For $k=10^{-2}$, a classical description should not be taken seriously. However, we include this mode nonetheless since its short wavelength renders it susceptible to the gradient instability according to the naive estimate above. These short-wavelength modes expose a general trend: in the approach to the bounce, the mode functions oscillate with a decreasing oscillation period and a growing amplitude. However, near the bounce, where all modes re-enter the horizon and where one might consequently expect either further oscillations or, due to the gradient instability, a strong growth, neither occurs. In fact, the curvature perturbations become nearly constant. We will describe and explain this interesting behavior in the next section. 

Our results can be compared with the analysis of Cai et al. \cite{Cai:2012va} (see also \cite{Cai:2013kja}). There, the authors studied an essentially identical model for the bounce \footnote{The differences between the two models are that: 1. Cai et al. consider higher-derivative terms with constant coefficients, while our higher-derivative terms are only turned on near the bounce. 2. Cai et al. have added a potential, which however does not play an important role near the bounce. In the regime of interest, i.e. near the bounce, both models are virtually identical.}, but found that even long-wavelength modes get amplified considerably--by as much as a (wavelength-independent) factor of $10^8$ according to Fig. 10 of \cite{Cai:2012va}. This is in contrast to our findings, which indicate that long-wavelength modes remain essentially unchanged. It is conceivable that the discrepancy is due to the fact that the authors of \cite{Cai:2012va} numerically solve the singular equation (\ref{eq:conformalR}) for $\mathcal{R}$. We will elaborate on this singularity momentarily. Before doing so, however, let us also compare our results with those of Xue et al. \cite{Xue2013}, who performed a non-perturbative numerical analysis of a bounce model with two scalar fields, with one of the scalars having a wrong-sign kinetic term. This field leads to a ghost, but classically one can still solve for the evolution across the bounce. Even though one cannot directly compare our bounce model to the one considered by Xue et al., it is interesting to note that in \cite{Xue2013} it was also found that curvature perturbations only get amplified by a small amount across the bounce\footnote{Other works of interest related to the question of the propagation of perturbations through a non-singular bounce include \cite{Peter:2002cn,Martin:2003sf,Allen:2004vz,Falciano:2008gt,Lilley:2011ag}.}. 

\begin{figure}[t!]
\begin{minipage}{\thirdWidthLeft} \flushleft
\includegraphics[width=\thirdWidthRight]{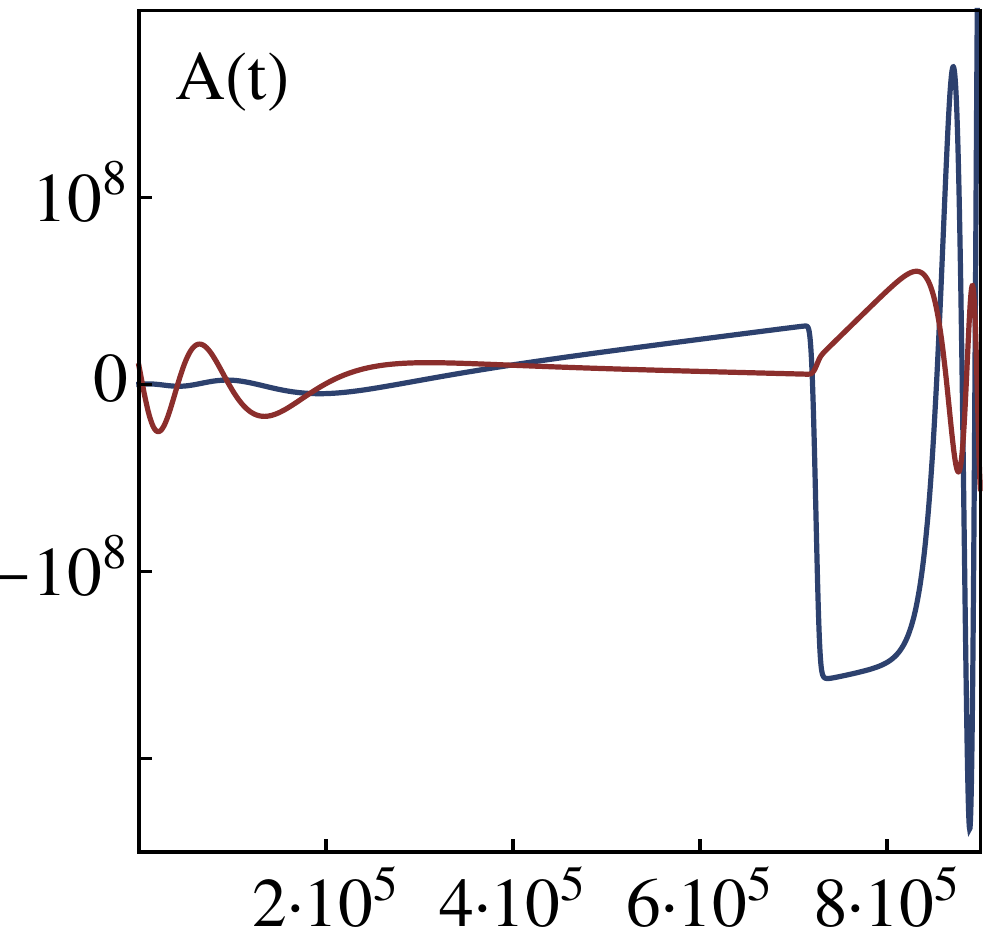}
\end{minipage}%
\begin{minipage}{\thirdWidthLeft} \flushleft
\includegraphics[width=\thirdWidthRight]{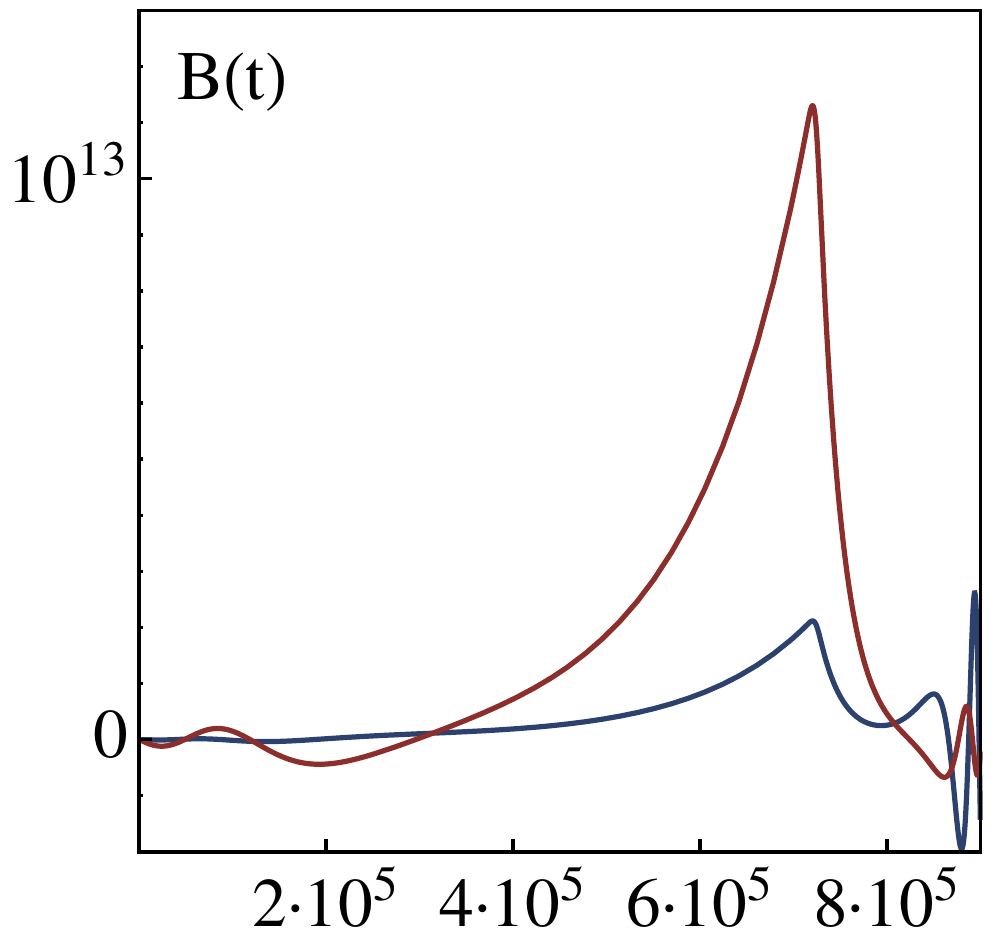}
\end{minipage}%
\begin{minipage}{\thirdWidthRight} \flushleft
\includegraphics[width=\thirdWidthRight]{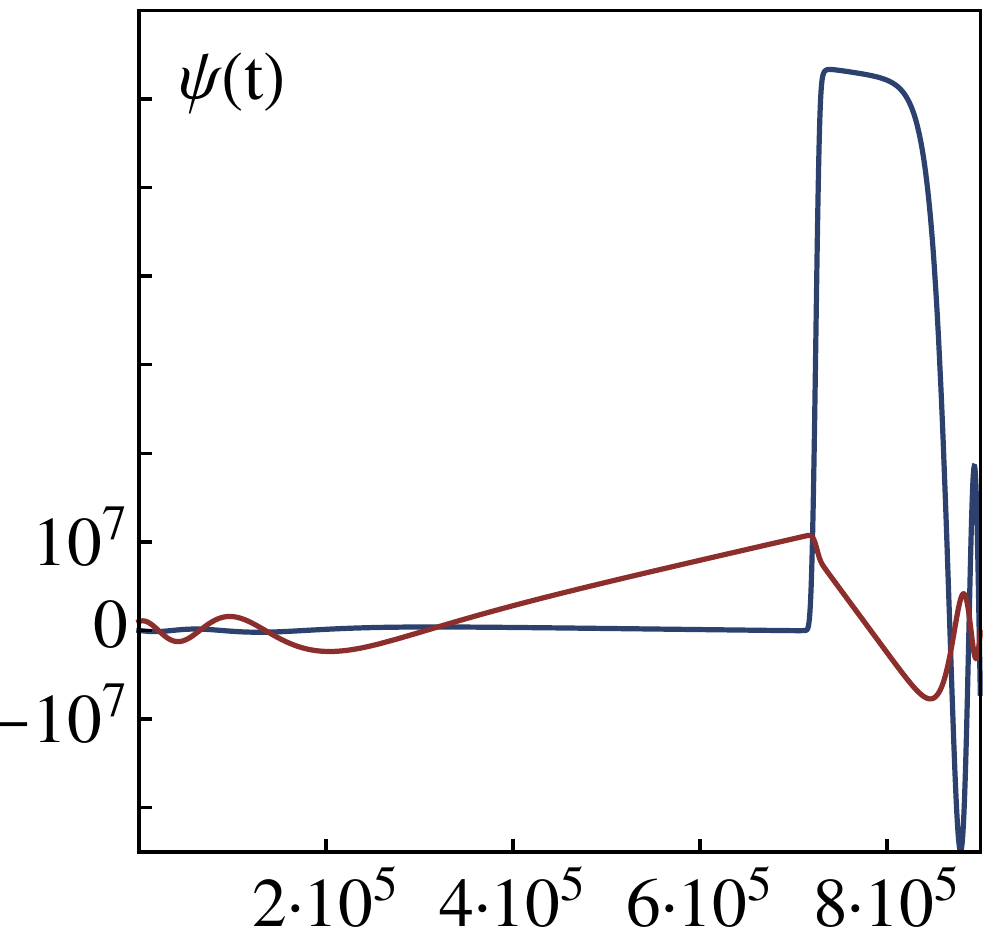}
\end{minipage} \vspace{.4cm}\\
\begin{minipage}{\thirdWidthLeft} \flushleft
\includegraphics[width=\thirdWidthRight]{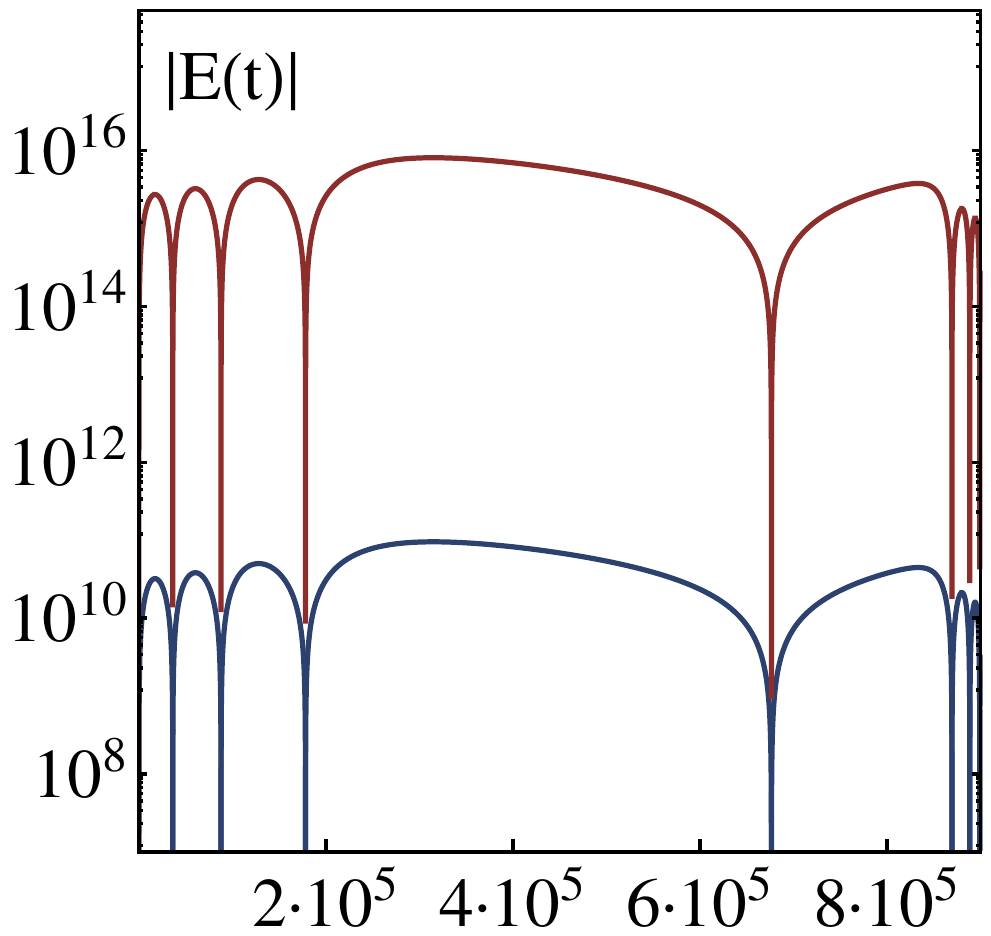}
\end{minipage}%
\begin{minipage}{\thirdWidthLeft} \flushleft
\includegraphics[width=\thirdWidthRight]{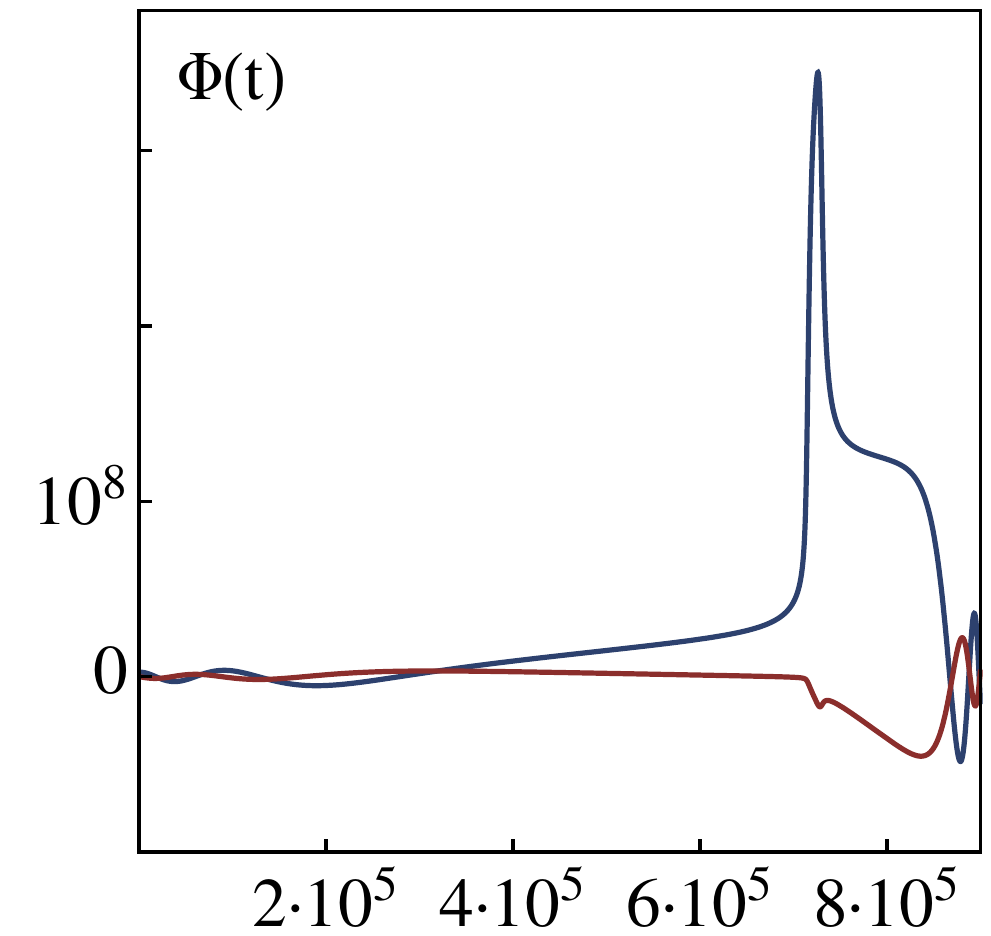}
\end{minipage}%
\begin{minipage}{\thirdWidthRight} \flushleft
\includegraphics[width=\thirdWidthRight]{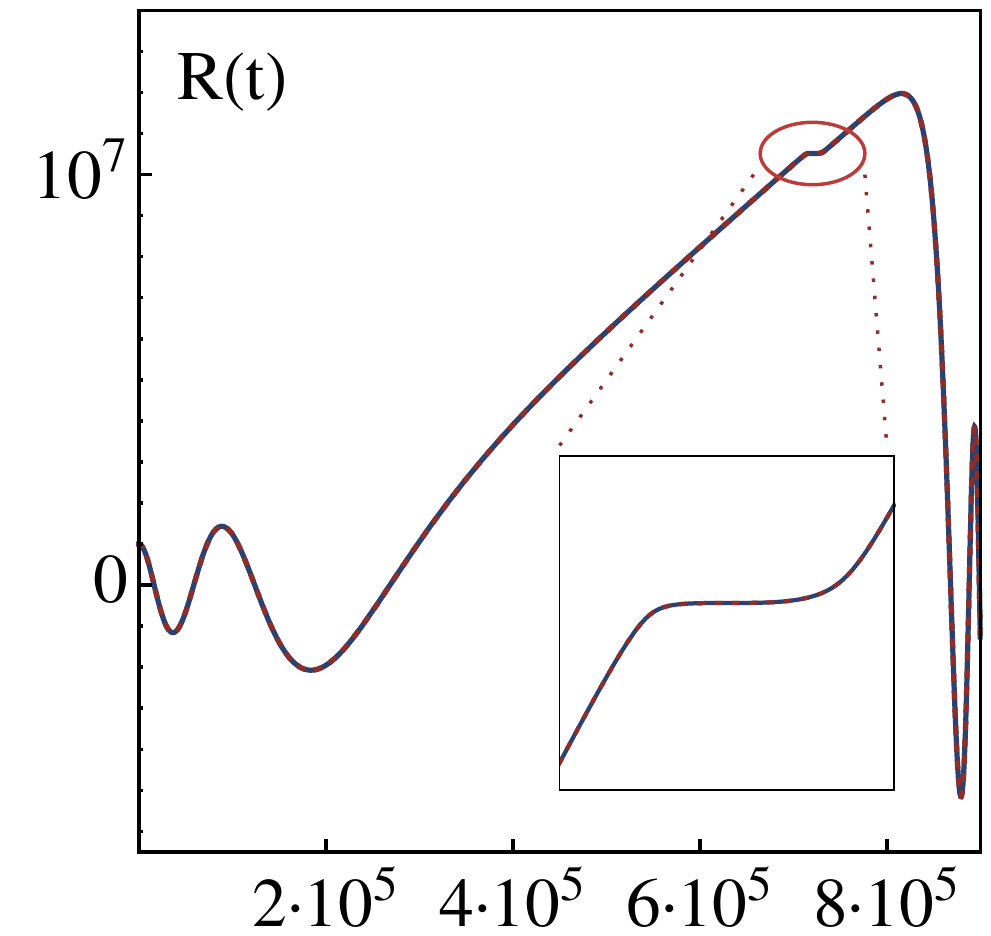}
\end{minipage}%
\caption{\label{fig:bounce6} Evolution of the perturbation variables in harmonic gauge with $k=10^{-4}$. The two colors represent the alternative gauge conditions $\mathbf{A}(t_0) = \mathbf{B}(t_0) = \psib(t_0) = \mathbf{E}(t_0) = 0$ (blue) and $\mathbf{A}(t_0) = 10^7$, $\mathbf{B}(t_0) = 10^{10}$, $\psib(t_0) = 10^3$, $\mathbf{E}(t_0) = 10^3$ (red). The profile of $\mathcal{R}(t)$ (right bottom panel) coincides in the two gauges, confirming the gauge invariance of (\ref{eq:finalPert1})--(\ref{eq:finalPert5}). The detail in the same plot shows that $\mathcal{R}$ evolves regularly across the bounce and, in particular, $R'$ vanishes when $ \mathcal{H} + \frac{1}{2} g \frac{ \phi ^{\prime 3}}{a ^6} = 0$ as discussed in Section \ref{section:singularities}.}
\end{figure}

As a check of our numerical investigation, we have verified that the result for the gauge--invariant variable $ \mathcal{R}$ is unchanged if different initial values for $ \mathbf{A}$, $ \mathbf{B}$, $\psib$ and $ \mathbf{E}$ are used (keeping in mind that one can set the initial values of $ \mathbf{A}$, $ \mathbf{B}$, $\psib$ and $ \mathbf{E}$ to any desired values using the residual gauge freedom of harmonic gauge discussed in section \ref{section:harmonic})--see Fig. \ref{fig:bounce6}. This also constitutes an explicit check of the gauge invariance of (\ref{eq:finalPert1})--(\ref{eq:finalPert5}). Note from Fig. \ref{fig:bounce6} that the individual metric perturbation functions have strongly gauge-dependent behavior, and that their individual evolution tends to be drastically different from that of the comoving curvature perturbation. 

We have solved the perturbation equations using both
Mathematica and C++ as programming languages. In the
latter case, we have implemented the fourth-order Runge-Kutta-Fehlberg
method with a fifth-order error estimator that controls the adaptive
stepsize. The mutual agreement up to high accuracy of these two
independent approaches makes us confident in the reliability of our results.

\subsection{Apparent singularities in solving for the curvature perturbation directly} \label{section:singularities}

Across the bounce, the gauge choice corresponding to constant mean curvature time slices does not produce well--behaved equations for linear perturbations. This happens because the mean curvature is non--monotonic in time, and constant-curvature slices are not necessarily spacelike in the presence of inhomogeneities. Similar issues occur for many standard gauge choices \cite{Xue2013}, and for this reason many gauge choices lead to linearized equations of motion that become singular close to/at the bounce. By contrast, as we have seen, harmonic gauge remains well--defined across the bounce. Nevertheless, it is informative to study the behavior of the equation of motion for $\mathcal{R}$ in a little more detail. As already discussed above, in conformal time it is given by 
\begin{equation} \label{eq:conformalR2}
\frac{ d ^2 \mathcal{R}}{d \tau ^2} + \frac{2}{z} \frac{ d z }{ d \tau} \frac{ d \mathcal{R}}{ d \tau} + c  _{s} ^2 k ^2 \mathcal{R} = 0 \;,
\end{equation}
where 
\begin{eqnarray}
z ^2 & = & \frac{1}{2} \frac{ a ^2 \phi ^{\prime 2} \mathcal{P}}{ \left( \mathcal{H} + \frac{1}{2} g( \phi) \frac{ \phi ^{\prime 3}}{a ^6} \right) ^2} \;,\label{eq:z22} \\
\mathcal{P} & = &  P_{,X} - 6 g( \phi) \mathcal{H} \frac{\phi'}{ a ^6} + \frac{3}{2} g ^{2}( \phi) \frac{ \phi ^{\prime 4}}{a ^{12}} + 2 g_{,\phi} \frac{ \phi ^{\prime 2}}{a ^6} + P_{,XX} \frac{ \phi ^{\prime 2}}{ a ^{6}} \;, \label{eq:calP2} \\
c_s ^2 & = &  \frac{1}{ \mathcal{P}} \left(P_{,X} + 2  g( \phi) \cH \frac{ \phi'}{ a ^6} - \frac{1}{2} g ^2( \phi) \frac{ \phi ^{\prime 4}}{a ^{12}} - 2 g( \phi) \frac{ \phi''}{a ^{6}} \right) \;.\label{eq:cs22}
\end{eqnarray}
The quadratic action for the perturbations takes the simple form
\begin{equation} \label{eq:action2R}
S_2  =  \int dt\, \frac{z ^2}{a ^{2}} \left( \frac{1}{2} \mathcal{R} ^{ \prime 2} - \frac{1}{2} c _{s} ^2 k ^2 a ^4 \mathcal{R} ^2 \right) \;.
\end{equation}

\begin{figure}[t]%
\begin{minipage}{\smallWidthLeft} \flushleft
\includegraphics[width=\smallWidthRight]{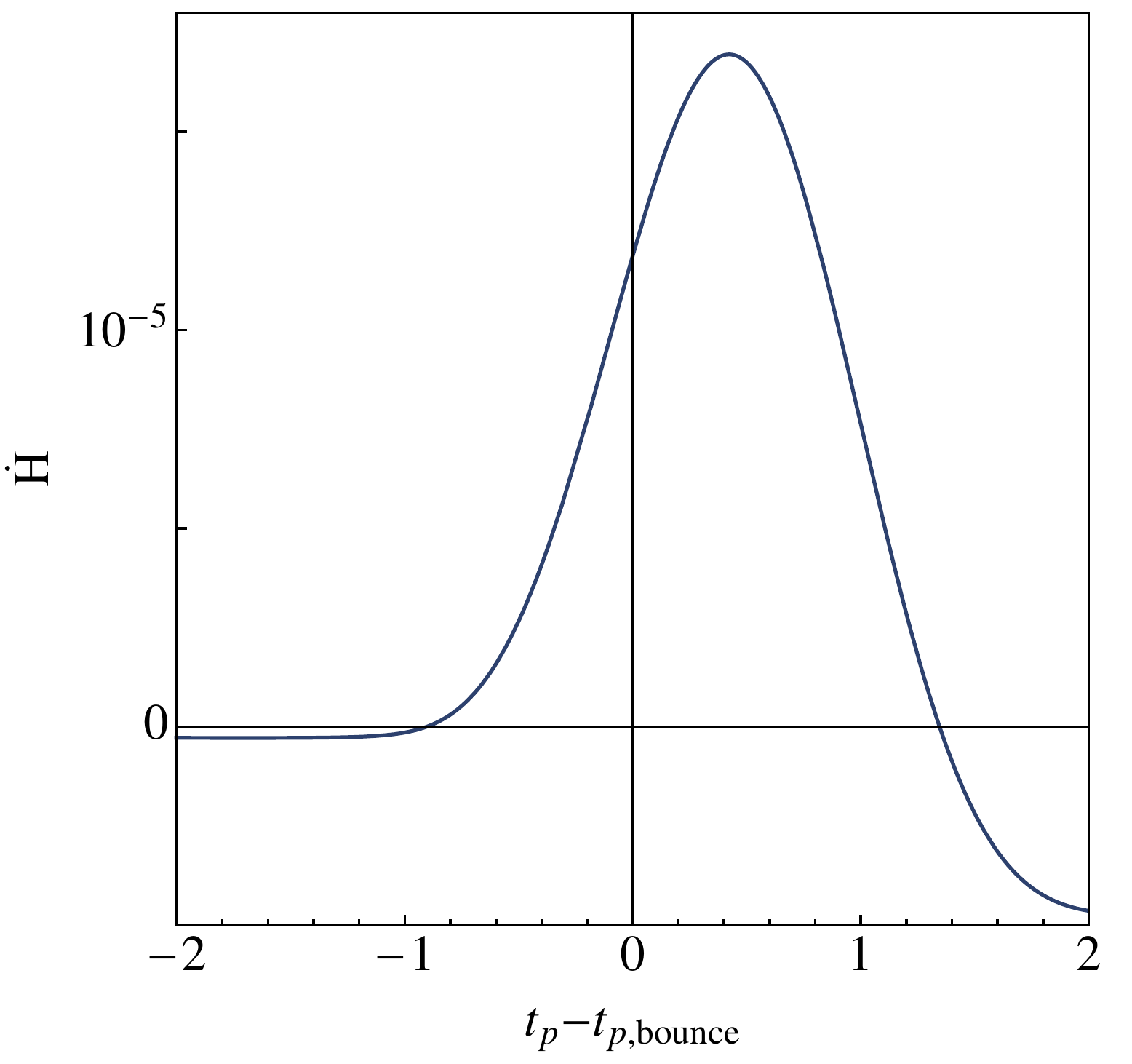}
\end{minipage}%
\begin{minipage}{\smallWidthRight} \flushleft
\includegraphics[width=\smallWidthRight]{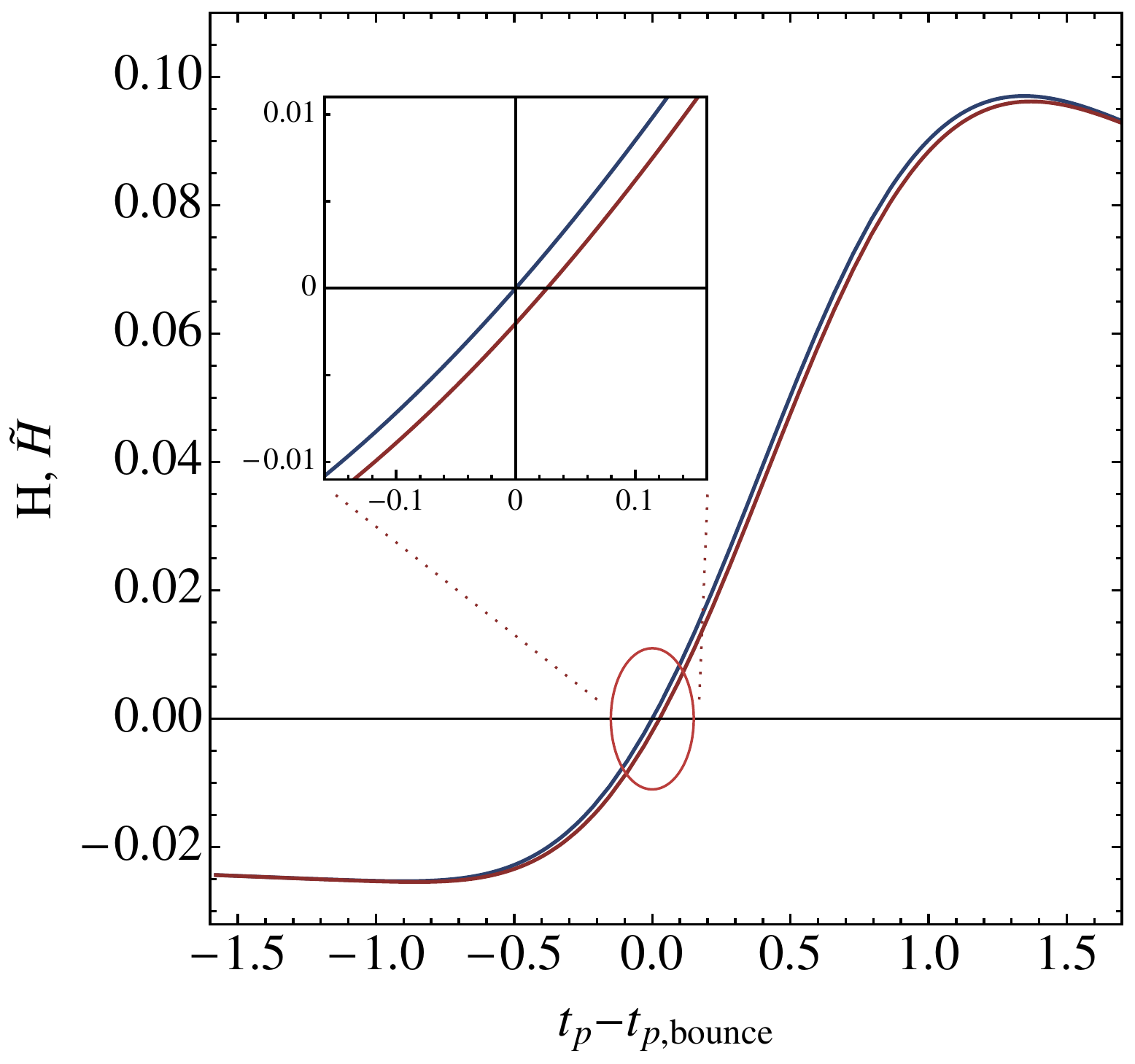}
\end{minipage}%
\caption{\label{fig:bounce7} The left panel shows the evolution of the rate of change of the Hubble parameter $\dot{H}\equiv (da/dt_p)/a$ in physical time - thus, positive $\dot{H}$ corresponds to a violation of the NEC, with the origin of the vertical axis being positioned at the moment of the bounce $H=0$. Note that the bounce is short - its duration is only a couple of time units, with the scale being given by the scale of the ghost condensate; for a ghost condensate occurring at the GUT scale (e.g. $t_b = 1/(10^{-2})^4 =10^8$), this would correspond to a bounce lasting ${\cal O}(t_b^{1/2}M_P) = {\cal O}(10^4)$ Planck times. The right panel depicts $H$ (blue, left curve) and $\tilde{\it H}\equiv H + \frac{1}{2}g \dot{\phi}^3$ (pink, right curve), again as functions of physical time. The moment at which the right curve crosses the horizontal axis corresponds to the singular time $t_s$ defined in Eq. (\ref{eq:nearBounce}). Note that this time is almost coincident with the moment of the bounce, and certainly well within the region of NEC violation.}
\end{figure}%

Close to the bounce, there is an apparent singularity as the denominator of $ z ^2$ goes through zero. This occurs at the time $t=t_s$ (or $\tau=\tau_s$ in conformal time) when
\begin{equation} \label{eq:nearBounce}
\mathcal{H} + \frac{1}{2} g \frac{ \phi ^{\prime 3}}{a ^6}  = 0 \quad \leftrightarrow \quad t = t_s \;.
\end{equation}
In our numerical example, the time $t_s$ is essentially identical to the time when $\mathcal{H}=0,$ due to the small numerical coefficient of the Galileon term--see Fig. \ref{fig:bounce7}. At this ``singular time'', 
there are two apparent singularities:
\begin{itemize}
\item in the equation of motion \eqref{eq:conformalR2}, the coefficient  of the $ d \mathcal{R}/ d \tau$ term diverges. This makes the perturbation equation singular except if $ d\mathcal{R} / d \tau$ vanishes when $ z ^2$ diverges.
\item in the action \eqref{eq:action2R}, even supposing that the combination $ z ^{2} \mathcal{R} ^{ \prime 2}$ remains finite at $ t = t_s$, the contribution from $ z ^2 \mathcal{R} ^2$ is divergent and non--integrable. 
\end{itemize}
In fact, a closer analysis reveals that both of these divergences are a manifestation of the ill--defined nature of $ \mathcal{R}$ as a genuine degree of freedom when $t = t_s$.

First, we show that $ \mathcal{R}' = 0$ at $t = t_s$. Inserting \eqref{eq:nearBounce} in \eqref{eq:finalPert4} we obtain
\begin{equation}
\psib' + \frac{\cH}{ \phi'} \Phib' - \frac{1}{2} \frac{ \phi'}{a ^6} \left( a ^6 P_{,X} - 3 g \cH \phi' + g_{,\phi} \phi ^{\prime 2} \right) \Phib = 0 \;, \quad (t = t_s)  \;
\end{equation}
and hence
\begin{eqnarray}
\mathcal{R'} & = & \psib' + \frac{\cH}{ \phi'} \Phib' + \left( \frac{ \cH'}{ \phi'} - \frac{\cH \phi''}{ \phi ^{\prime 2}} \right) \Phib \nonumber \\
& = & \frac{1}{ \phi'}\left( \cH'- \frac{\cH \phi''}{ \phi ^{\prime}} + \frac{1}{2} P_{,X} \phi ^{\prime 2} - \frac{3}{2} g \cH \frac{ \phi ^{\prime 3}}{a ^6} + \frac{1}{2} g_{,\phi} \frac{ \phi ^{\prime 4}}{a ^6}  \right) \Phib \;,  \quad (t = t_s) \label{eq:rprimebounce}\;.
\end{eqnarray}
Summing the two background equations \eqref{bgharm1} and \eqref{bgharm2} one gets
\begin{equation}
\cH' + \frac{1}{2} g \frac{ \phi ^{\prime 3}}{a ^6} \frac{ \phi''}{ \phi'}= - \frac{1}{2} P_{,X} \phi ^{\prime 2} + 3 \cH \left(\cH + g \frac{ \phi ^{\prime 3}}{a ^6} \right)   - \frac{1}{2} g_{,\phi} \frac{ \phi ^{\prime 4}}{a ^6} \;.
\end{equation}
Using \eqref{eq:nearBounce} simplifies this expression at the singular time to
\begin{equation}
\mathcal{H}' - \frac{ \cH \phi''}{ \phi'} = - \frac{1}{2} P_{,X} \phi ^{\prime 2} + \frac{3}{2} g \cH \frac{ \phi^{\prime 3}}{a ^6} - \frac{1}{2} g_{,\phi} \frac{ \phi ^{\prime 4}}{a ^6} \;,  \quad (t = t_s) \;.
\end{equation}
Hence, the bracket in \eqref{eq:rprimebounce} vanishes and so does $ \mathcal{R}'$. Let us highlight this result:
\begin{equation}
\mathcal{R}^\prime = 0 \;, \quad \forall k \;, \quad (t=t_s) \;.
\end{equation}
Thus, we have now discovered a proof that the flattening out of the perturbation modes near the bounce, observed in the numerical evaluation above, must necessarily occur for all modes precisely in order to avoid a singular solution of Eq. (\ref{eq:conformalR2}). Note that this result is independent of whether or not a gradient instability is present. One can also see this result by directly considering the general solution (now in conformal time) of \eqref{eq:conformalR2} near $ \tau = \tau_s$. First, note that near the singular time one has
\begin{equation}
\frac{z'}{z} \sim -\frac{1}{( \tau- \tau_s)} \;,
\end{equation}
while $c_s$ approaches some finite constant $ \bar{c}_s$. Equation \eqref{eq:conformalR2} can be re--written as
\begin{equation}
\frac{d}{d \tau}
\left(
\begin{array}{c} \mathcal{R} \\ \frac{d \mathcal{R}}{ d \tau} \end{array}
 \right) = \left(
 \begin{array}{cc}
 0 & 1 \\
 - c_s ^2 k ^2 & - \frac{2}{z} \frac{ d z}{ d \tau} \end{array}
  \right) \left(
\begin{array}{c} \mathcal{R} \\ \frac{d \mathcal{R}}{ d \tau}  \end{array}
 \right) \;.
 \end{equation}
So the system is singular at $ \tau  = \tau_s.$ The general solution of \eqref{eq:conformalR2} near $ \tau = \tau_s$ is given by
 \begin{equation}
 \mathcal{ R} = \alpha  \left(1 - \frac{1}{2} \bar{c}_s ^2 k ^2 ( \tau - \tau_s) ^2 + \ldots \right) + \beta ( \tau - \tau_s) ^3 \;,
 \end{equation}
where $\alpha,\beta$ are constants. Thus, once again, we find that $ \mathcal{R}' = 0$ at $ \tau = \tau_s$. The singularity in the equation for $ \mathcal{R}$ forces this relation to hold. This ensures that, although the equation is singular, its solution is not.

The result above shows that the combination $ z ^{2} \mathcal{R} ^{ \prime 2}$ remains finite as $ t \rightarrow t_s$. So, what remains to be done is to explain the origin of the divergence of the term $ z ^2 \mathcal{R} ^2$ in the action \eqref{eq:action2R}. As we pointed out above, the basic perturbation variables $ \mathbf{A}$, $ \mathbf{B}$, $\psib$, $ \mathbf{E}$ and $\Phib$ evolve smoothly across the bounce. Therefore, since their coefficients in the unconstrained, non--gauge fixed quadratic action have no divergence, the latter should be perfectly finite when expressed in the original variables. In order to understand the origin of the divergence, one must reconsider the derivation of \eqref{eq:action2R} in the $\Phib = 0$, $\mathbf{E} = 0$ gauge, which is well--defined across the bounce (see \eqref{eq:gt4}). In this gauge, $ \mathcal{R} \equiv \psib$ and \eqref{eq:finalPert4} imply that 
\begin{equation}
\left( \cH + \frac{1}{2} g\frac{ \phi ^{\prime 3}}{a ^6} \right) \mathbf{A} + \mathcal{R} ' = 0 \;.
\end{equation}
This constraint equation can be used to replace $ \mathbf{A}$ in the quadratic action by its expression in terms of $ \mathcal{R}'$ and $ \tilde{ \cH},$ where we define
\begin{equation}
 \tilde{ \cH} \equiv \cH + \frac{1}{2} g \frac{ \phi ^{\prime 3}}{a ^6} \;.
 \end{equation}
A term proportional to $ \mathcal{R} ^2$ is obtained when integrating by parts a term of the form
\begin{equation}
S_2 \sim \int dt \mathbf{A} \psib  \sim - \int dt \frac{\mathcal{R} \mathcal{R}'}{ \tilde{ \cH}}\;.
\end{equation}
Notice that this is perfectly finite, since $ \mathcal{R}'$ and $ \tilde{\cH}$ vanish simultaneously and at the same rate. However, when we integrate by parts we obtain
\begin{equation}
S_2 \sim  \frac{1}{2} \int dt \frac{ \tilde{\cH}'}{ \tilde{\cH} ^2} \mathcal{R} ^2 + \textrm{bulk term} \;.
\end{equation}
The bulk term is now divergent in the same way as the term we considered initially. However, this divergence is only the counterpart of the divergent boundary term of the integration by parts. Analogous potential divergences occur repeatedly in the derivation of the quadratic action for the comoving curvature perturbation. However, if one carefully keeps all boundary terms--with the subtlety that the ``boundary'' here occurs right in the middle of the dynamical evolution--divergences are avoided.

The discussion in the present section suggests that one can choose any gauge that one likes to calculate the transfer of perturbations across a bounce, as long as one is careful in dealing with singularities. However, picking a gauge which is entirely non-singular, such as the harmonic gauge employed in this paper, is likely to make the calculation much simpler and clearer. 

\section{Discussion}

Our results imply that in ekpyrotic/cyclic models of the universe, where the perturbations of interest are much larger than the horizon size at the onset of the bounce, the perturbations generated during an ekpyrotic contraction phase carry over unchanged into the expanding phase. Thus, if nearly scale-invariant curvature perturbations are produced (e.g. via the entropic mechanism \cite{Fertig:2013kwa,Ijjas:2014fja}), then they will re-enter the horizon during the expanding phase with the same amplitude and spectrum that they acquired during contraction. Our results thus explicitly demonstrate that there exist well-motivated models of non-singular bounces which allow one to trace cosmological perturbations unambiguously all the way from their generation during a contracting phase up to the present time.

Our results also have implications for the matter bounce scenario \cite{Cai:2013kja}. In this scenario, during a phase of pressure-free contraction (matter dominated contraction) scale-invariant scalar and tensor perturbations are generated with comparable amplitudes. Based on our current results, it seems reasonable to speculate that long-wavelength gravitational waves also get transferred unchanged through the bounce. In this case, the amplitude of the gravitational wave modes emerging in the expanding phase is typically too large compared to the scalar fluctuations, since a tensor-to-scalar ratio of $r \gtrsim 10$ is typically obtained. It was hoped that the bounce could amplify the scalar perturbations relative to the tensors \cite{Cai:2013kja}. However, as we have demonstrated, this is not the case, at least in bounce models of the ghost condensate/Galileon type. Thus, an interesting question for future research is whether the scalar fluctuations can be boosted relative to the tensors by some other mechanism. 

In our analysis, we discovered a number of interesting aspects in the behavior of short-wavelength modes, i.e. of modes whose wavelength is comparable or even smaller than the horizon size at the onset of the bounce. The classical treatment of the present paper is not fully adequate, however, for such short modes. It would clearly be interesting to perform a quantum-field-theoretic calculation of the evolution of such short modes in a classical non-singular bounce spacetime. In particular, one might imagine a scenario where a bounce is followed by a period of inflation, in which case there may be observable consequences (see e.g. \cite{Piao:2003zm,Liu:2013kea}). We hope to report on such an investigation in the future.


\acknowledgments

We would like to thank Robert Brandenberger, Yi-Fu Cai, Paul Steinhardt and BingKan Xue for useful correspondence and discussions. L.B., M.K. and J.L.L. gratefully acknowledge the support of the European Research Council via the Starting Grant Nr. 256994 ``StringCosmOS''. B.A.O. is  supported in part by the DOE under contract No. DE-SC0007901 and by the NSF under grant No. 1001296.


\appendix
\section{Derivation of the linearized equations of motion in harmonic gauge}

We consider the following model for the bounce:
\begin{eqnarray}
\mathcal{L} & = & \sqrt{-g} \left(\frac{R}{2} + P(X, \phi) + g( \phi) X\,  \Box \phi \right) \;, \\
X & \equiv & - \frac{1}{2} g ^{ \mu \nu} \partial _{\mu} \phi\, \partial _{\nu} \phi \;.
\end{eqnarray}
The Einstein and scalar field equations read (see e.g. \cite{Gao2011})
\begin{eqnarray}
&&G_{ \mu \nu}   =   T_{ \mu \nu} = T _{ \mu \nu} ^{P} + T _{ \mu \nu} ^{g} \;, \\
&& \mathcal{E} ^{P} + \mathcal{E} ^{g}  =  0 \;,
\end{eqnarray}
where
\begin{eqnarray}
T_{ \mu \nu} ^{P} & = & P\, g _{\mu\nu} + P_{,X} \partial _{\mu} \phi\, \partial _{\nu} \phi \;,\\
T_{ \mu \nu} ^{g} & = & - g _{\mu\nu} [g( \phi)  \nabla _{\mu} \phi \nabla ^{\mu} X - 2 g_{, \phi} X ^2] + \partial _{\mu} \phi\, \partial _{\nu} \phi [g( \phi) \Box \phi + 2 g_{,\phi} X] \nonumber \\ && + g( \phi) \partial _{ \mu} \phi\, \partial _{ \nu} X + g( \phi) \partial _{ \nu} \phi\, \partial _{ \mu} X \;, 
\end{eqnarray}
and
\begin{eqnarray}
\mathcal{ E} ^{P} & = & - P_{,XX} \nabla _{\mu} \phi \, \nabla _{\nu} \phi\, \nabla ^{\mu} \nabla ^{\nu} \phi - 2 X P_{, X \phi} + P_{, X}\, \Box \phi + P_{, \phi} \;, \\
\mathcal{E} ^{g} & = & g( \phi) \Big( (\Box \phi) ^2 - \nabla _{\mu} \nabla _{\nu} \phi\, \nabla ^{\mu} \nabla ^{\nu} \phi - R_{ \mu \nu} \nabla ^{\mu} \phi\, \nabla ^{\nu} \phi \Big) \nonumber\\&& - 2 g_{, \phi} \nabla _{\mu} \phi\, \nabla _{\nu}  \phi \, \nabla ^{\mu} \nabla ^{\nu} \phi - 2 g_{, \phi\phi} X ^2 \;.
\end{eqnarray}
In harmonic coordinates, the metric element and scalar field ansatz we are using are given by
\begin{eqnarray}
ds ^2 & = & - a ^{6}(1 + 2 \mathbf{A}) \d t ^2 + 2 a ^{4} \mathbf{B}_{,i}\, \d t\, \d x ^{i} + a ^2(t) \Big[ \left(1- 2 \psib \right) \delta _{ij} + 2 \mathbf{E}_{, ij} \Big] \d x ^{i} \d x ^{j} \;,\\
\phi & = & \phi(t) + \Phib(t, x) \;.
\end{eqnarray}
Accordingly, we find the following variations of the components of the Einstein tensor:
\begin{eqnarray}
\delta G \indices{^t _t} & = & \frac{2}{a ^{6}} \left\{3 \cH\left( \psib' + \cH \mathbf{A} \right) - \nabla^2 \left[a ^{4} \boldsymbol \psi + \cH ( \mathbf{E}' - a ^{2} \mathbf{B} ) \right] \right\} \;, \\
\delta G \indices{ ^{t} _{i}} & = & - \frac{2}{a ^{6}} \left\{ \psib'+ \cH \mathbf{A} \right\}_{,i} \;,\\
\delta G \indices{^i _j} & = & \delta^i_{j} \left\{ \frac{2 \cH}{a ^{6}} \mathbf{A}' + \frac{1}{a ^{6}} \left[4 \cH' - 6 \cH ^2 + a ^{4} \nabla^2 \right] \mathbf{A} + \frac{1}{a ^{4}} \nabla ^2 \mathbf{B}' + 2\frac{\cH}{a ^{4}} \nabla ^2 \mathbf{B}\right. \nonumber\\&& \left.
+ \frac{2}{a ^{6}} \psib'' - \frac{1}{a ^{2}} \nabla ^2 \psib - \frac{1}{a ^{6}} \nabla ^2 \mathbf{E}''  \right\} 
+ \frac{\delta^{ik}}{a ^2}\left\{- \mathbf{A} - \frac{1}{a ^{2}} \mathbf{B}' - 2 \frac{ \cH}{a ^{2}} \mathbf{B} + \boldsymbol \psi + \frac{1}{a ^{4}} \mathbf{E}'' \right\}_{,kj} \;,\\
\delta R & = & - \frac{6 \cH}{a ^{6}} \mathbf{A}' + \frac{2}{a ^{6}} \left[ - 6 \cH' + 6 \cH ^2 - a ^{4} \nabla^2 \right] \mathbf{A} - \frac{2}{a ^{4}} \nabla ^2 \mathbf{B}' - \frac{6 \cH}{a ^{4}} \nabla^2 \mathbf{B} \nonumber\\&&
- \frac{6}{a ^{6}} \psib'' - \frac{6 \cH}{a ^{6}} \psib' + \frac{4}{a ^{2}} \nabla ^{2} \psib 
+ \frac{2}{a ^{6}} \nabla ^2 \mathbf{E}'' + \frac{2 \cH}{a ^{6}} \nabla ^2 \mathbf{E}'  \;,
 \end{eqnarray}
while the perturbed stress--energy tensor is given by
\begin{eqnarray}
(\delta T^P) \indices{^ t _ t}  & = & \left( \frac{P_{,X} \phi ^{ \prime 2}}{a ^{6}} + P_{, X X} \frac{ \phi ^{ \prime 4}}{a ^{12}} \right) \mathbf{ A} - \frac{1}{a ^{6}} \left( P_{,X} \phi' + P_{, XX} \frac{ \phi ^{ \prime 3}}{a ^{6}} \right) \Phib' \nonumber\\&& + \frac{1}{a ^{6}} \left( a ^{6} P_{, \phi} - P_{, X \phi} \phi ^{ \prime 2} \right) \Phib \;, \\
(\delta T^P) \indices{^ t _i} & = & - \frac{P_{,X} \phi'}{a ^{6}} \Phib_{,i} \;,\\
(\delta T^P) \indices{^i _t} & = & \frac{P_{,X} \phi ^{ \prime 2}}{a ^{4}} \delta^{ik}\mathbf{B}_{, k} + \frac{P_{,X} \phi'}{ a ^{2}} \delta^{ik}\Phib_{,k} \;, \\
(\delta T^P) \indices{ ^{i} _{j}} & = & \delta ^i_{j} \left\{ - \frac{P_{,X} \phi ^{ \prime 2}}{a ^{6}} \mathbf{A} +\frac{P_{,X} \phi'}{ a ^{6}} \Phib' + P_{, \phi}  \Phib \right\} \;.
\end{eqnarray}
\begin{eqnarray}
(\delta T^g) \indices{ ^{t}_t } & = & - 2 \frac{ \phi ^{ \prime 3}}{ a ^{12}} \left(6 g( \phi) \mathcal{H} - g_{,\phi} \phi' \right) \mathbf{A} - \frac{g( \phi) \phi ^{ \prime 3} \nabla ^2}{a ^{10}} \mathbf{B} - 3 \frac{g( \phi) \phi ^{ \prime 3}}{a ^{12}} \psib' + \frac{g( \phi) \phi ^{ \prime 3} \nabla ^2 }{a ^{12}} \mathbf{E}' \nonumber\\&&
+ \frac{ \phi'}{a ^{12}} \left[ 9 g( \phi) \cH \phi' - 2 g_{,\phi} \phi ^{\prime 2} \right] \Phib' \nonumber\\&&
+ \frac{ \phi'}{ a ^{12}} \left[ - g( \phi) a ^{4}\phi' \nabla ^2 + 3 g_{,\phi} \cH \phi ^{ \prime 2} - \frac{1}{2} g_{,\phi\phi} \phi ^{ \prime 3} \right] \Phib \;, \\
(\delta T^g) \indices{ ^{t} _{i}} & = & \left\{ \frac{g( \phi) \phi ^{ \prime 3}}{a ^{12}} \mathbf{A} 
- \frac{1}{a ^{12}}g( \phi) \phi ^{\prime 2} \Phib' - \frac{ \phi'}{a ^{12}} \left[ -3 g( \phi) \cH \phi' + g_{,\phi} \phi ^{ \prime 2} \right] \Phib \right\}_{, i} \;, \\
(\delta T^g) \indices{ ^{i} _{t}} & = & \delta^{ik} \left\{ - \frac{g( \phi) \phi ^{ \prime 3}}{a ^{8}} \mathbf{A} + \frac{ \phi ^{ \prime 2}}{a ^{10}} \left[ g( \phi) \left( \phi'' - 6 \cH \phi' \right) + g_{,\phi} \phi ^{ \prime 2} \right] \mathbf{B} \right. \nonumber\\&& \left.
+ \frac{1}{a ^{8}} g( \phi) \phi ^{\prime 2} \Phib'+ \frac{ \phi'}{a ^{8}} \left[ -3 g( \phi)\cH \phi'  + g_{,\phi} \phi ^{ \prime 2} \right] \Phib \right\}_{, k} \;, \\
(\delta T^g) \indices{ ^{i} _{j}} & = & \delta ^i_{j} \left\{ - \frac{1}{a ^{12}} g( \phi) \phi ^{\prime 3} \mathbf{A}' - \frac{ \phi ^{ \prime 2}}{a ^{12}} \left[ g( \phi) \left( - 12\cH \phi' + 4 \phi'' \right) + 2 g_{,\phi} \phi ^{ \prime 2} \right] \mathbf{ A} \right. \nonumber\\&& \left.
+ \frac{1}{a ^{12}} g( \phi) \phi ^{\prime 2} \Phib'' + \frac{ \phi'}{a ^{12}} \Big[ g( \phi) \left(-9 \cH \phi' + 2 \phi'' \right) + 2 g_{,\phi} \phi ^{\prime 2} \Big] \Phib' \right. \nonumber \\ && \left.
+ \frac{ \phi'}{a ^{12}} \left[ g_{,\phi}\phi' \left(- 3 \cH \phi' + \phi'' \right) + \frac{1}{2} g_{,\phi\phi} \phi ^{ \prime 3}\right]  \Phib \right\} \;.
\end{eqnarray}
The linearized scalar field equation follows from the linearized Einstein equations. For completeness we derive it here:
\begin{equation}
 \delta \mathcal{E}  = \delta \mathcal{E} ^{P} + \delta \mathcal{E} ^{g} \;,
 \end{equation}
with
\begin{eqnarray}
\delta \mathcal{E} ^{P} & = & \left\{ \frac{1}{a ^{6}} P_{,X} \phi' + \frac{1}{a ^{12}} P_{,XX} \phi ^{\prime 3} \right\} \mathbf{A}' + \left\{ \frac{2}{a ^{6}} P_{,X} \phi'' + P_{,X \phi}\frac{ \phi ^{\prime 2}}{ a ^{6}} +P_{,XX} \frac{ \phi ^{\prime 2}}{a ^{12}} \left[ 5 \phi'' - 12 \cH \phi' \right] \right. \nonumber\\&& \left.
+ P_{,XX \phi} \frac{ \phi ^{\prime 4}}{a ^{12}} + P_{,XXX} \frac{ \phi ^{\prime 4}}{a ^{18}} \left[ \phi'' - 3 \cH \phi' \right] \right\} \mathbf{A} 
+ P_{,X} \frac{ \phi' \nabla ^{2}}{ a ^{4}} \mathbf{B} + 3 P_{,X} \frac{ \phi' }{a ^{6}} \psib'-P_{,X} \frac{ \phi' \nabla ^{2}}{a ^{6}} \mathbf{E}'\nonumber\\&&
+ \left\{ - \frac{1}{a ^{6}} P_{,X} - \frac{1}{a ^{12}} P_{,XX} \phi ^{\prime 2} \right\} \Phib''  + \left\{ - \frac{1}{a ^{6}}P_{,X \phi} \phi' - P_{,XX} \frac{ \phi'}{ a ^{12}} \left(3 \phi'' - 9 \cH \phi' \right) \right. \nonumber \\ && \left. - \frac{1}{a ^{12}} P_{,XX \phi} \phi ^{\prime 3}  - P_{,XXX} \frac{ \phi ^{\prime 3}}{a ^{18}} \left( \phi'' - 3 \cH \phi' \right) \right\} \Phib' \nonumber \\ &&
+ \left\{ \frac{1}{a ^2} P_{,X} \nabla ^2  + P_{, \phi \phi} - \frac{P_{,X \phi}}{a ^{6}} \phi''  
- P_{, X \phi \phi} \frac{ \phi ^{ \prime 2}}{a ^{6}}- P_{,XX \phi}\frac{ \phi ^{ \prime 2}}{a ^{12}} \left[ \phi'' - 3 \cH \phi' \right] \right\} \Phib \;,\\
\delta \mathcal{E} ^{g} & = & \frac{1}{a ^{12}}\Big\{ -9 g( \phi) \cH \phi ^{\prime 2} + 2 g_{,\phi}\phi ^{\prime 3} \Big\} \mathbf{A}' +
\frac{1}{a ^{12}} \Big\{ g( \phi)  \phi' \left[ -24 \cH \phi''+ 72 \cH ^{2} \phi'  - 12 \cH' \phi' - a ^{4} \phi' \nabla ^{2} \right]   \nonumber\\&& 
+ 2g_{,\phi} \phi ^{\prime 2} \left[ 4 \phi'' - 12 H \phi' \right] + 2 g_{,\phi\phi}\phi ^{ \prime 4} \Big\} \mathbf{A} \nonumber\\&&
- \frac{1}{a ^{10}} g( \phi) \phi ^{\prime 2} \nabla ^2 \mathbf{B}' 
- g( \phi) \frac{ \phi'}{ a ^{10}} \left[  2 \phi'' -  \cH \phi' \right] \nabla ^2 \mathbf{B} - \frac{3}{a ^{12}} \phi ^{\prime 2} \psib'' - 3g( \phi) \frac{ \phi'}{a ^{12}} \left[2\phi'' - 3 \cH \phi' \right] \psib' \nonumber\\&&
+ \frac{1}{a ^{12}} g( \phi) \phi ^{\prime 2} \nabla ^2 \mathbf{E}'' + g( \phi) \frac{ \phi'}{a ^{12}} \left[ 2 \phi'' - 3\phi' \cH \right] \nabla ^2 \mathbf{E}' \nonumber\\&&
+ \frac{1}{a ^{12}} \Big\{ 6 g( \phi) \cH \phi' - 2g_{,\phi} \phi ^{\prime 2} \Big\} \Phib'' \nonumber \\
&& + \frac{1}{a ^{12}} \Big\{ 2 g( \phi) \left(3 \cH \phi'' - 18 \cH ^2 \phi' + 3 \phi' \cH' \right) - g_{,\phi} \phi' (4 \phi'' - 18 \cH \phi') - 2 g_{,\phi\phi} \phi ^{\prime 3} \Big\} \Phib' \nonumber \\&&
+ \frac{1}{a ^{12}} \Big\{ 2g( \phi) \left[- a ^{4} \left( \phi'' - \cH \phi' \right) \nabla ^2 \right] 
-g_{,\phi} \phi'\left[  - 6\cH \phi'' + 18 \cH ^2 \phi' - 3\cH' \phi' \right] \nonumber\\&&
- 2 g_{,\phi\phi}\phi ^{ \prime 2} \left[ \phi'' - 3 \cH \phi' \right] - \frac{1}{2} g_{,\phi\phi\phi} \phi ^{ \prime 4} \Big\} \Phib \;.
\end{eqnarray}
Using the results above, one can obtain the linearized field equations:
\begin{eqnarray}
\mathbf{0} & \mathbf{=} & \mathbf{\frac{a ^{6}}{2} \Big( \delta G \indices{ ^{t} _{t}} - \delta T \indices{ ^{t} _{t}}\Big)}  : \nonumber\\ 
\label{t-t} 0 & = & 
\Big( 3 \cH ^2 - \frac{1}{2} P_{, X} \phi ^{\prime 2} - \frac{1}{2} P_{,XX} \frac{ \phi ^{\prime 4}}{a ^{6}} + 6 g( \phi) \cH \frac{ \phi ^{\prime 3}}{ a ^{6}} - g_{,\phi} \frac{ \phi^{\prime 4}}{a ^{6}} \Big) \mathbf{A} - k ^2\Big(a ^2 \cH +\frac{1}{2} g( \phi) \frac{ \phi ^{\prime 3}}{a ^{4}} \Big) \mathbf{B}\nonumber\\&& + 3 \left(\cH + \frac{1}{2} g( \phi) \frac{ \phi ^{\prime 3}}{a ^6} \right) \psib'+ k ^2 a ^4  \psib + k ^2 \left( \cH + \frac{1}{2} g( \phi) \frac{ \phi ^{\prime 3}}{a ^6} \right) \mathbf{E}' \nonumber\\&&
+ \frac{1}{2} \left(P_{,X} \phi' + P_{,XX} \frac{ \phi ^{\prime 3}}{a ^{6}} - 9 g( \phi) \cH \frac{ \phi ^{\prime 2}}{a ^6} + 2 g_{,\phi} \frac{ \phi ^{\prime 3}}{a ^6} \right) \Phib' \nonumber\\&&
- \frac{1}{2} \left( a ^6 P_{, \phi} - P_{, X \phi} \phi ^{\prime 2} + k ^2 g( \phi) \frac{ \phi ^{\prime 2}}{a ^2} + 3 g_{,\phi} \cH \frac{ \phi^{\prime 3}}{a ^6} - \frac{1}{2} g_{,\phi\phi} \frac{ \phi ^{\prime 4}}{a ^6} \right) \Phib \;,\\
\mathbf{0} & \mathbf{=} & \mathbf{- \frac{a ^{6}}{2} \Big( \delta G \indices{ ^{t} _{i}} - \delta T \indices{ ^{t} _{i}}\Big)}  : \nonumber\\
\label{t-i} 0 & = & \left(\cH + \frac{1}{2} g( \phi) \frac{ \phi ^{\prime 3}}{a ^6} \right) \mathbf{A} +  \psib'   - \frac{1}{2} g( \phi) \frac{ \phi ^{\prime 2}}{a ^{6}} \Phib' 
- \frac{1}{2} \frac{ \phi'}{a ^6} \left( a ^6 P_{,X} - 3 g( \phi) \cH \phi' + g_{,\phi}\phi ^{\prime 2} \right) \Phib \;,\\
\mathbf{0} & \mathbf{=} & \mathbf{a ^{6} \Big( \delta G \indices{ ^{i} _{j}} - \delta T \indices{ ^{i} _{j}}\Big) ^{T}}  : \nonumber\\
0 & = & 
\left( 2 \cH + g( \phi) \frac{ \phi ^{\prime 3}}{a ^6} \right) \mathbf{A}' + \Big[ 4 \cH' - 6 \cH ^2 - k ^2a ^4 + P_{,X} \phi ^{\prime 2} + \frac{ \phi ^{\prime 2}}{a ^6} \left(g( \phi) (-12 H \phi' + 4 \phi'') + 2 g_{,\phi} \phi ^{\prime 2} \right) \Big] \mathbf{A} \nonumber\\&&
- k ^2 a ^2 \mathbf{B}' -2 k ^2a ^2 \cH \mathbf{B} + 2 \psib'' + k ^2 a ^4 \psib + k ^2 \mathbf{E}''- g( \phi) \frac{ \phi ^{\prime 2}}{a ^6} \Phib'' \nonumber\\&& - \left(P_{,X} \phi' + g( \phi)\frac{ \phi'}{a ^{6}} \left(-9H \phi' + 2 \phi'' \right) + 2 g_{,\phi} \frac{ \phi ^{\prime 3}}{a ^{6}} \right) \Phib' \nonumber\\&& - \left(a ^6 P_{, \phi} + g_{,\phi} \frac{ \phi ^{\prime 2}}{a ^{6}} \left(-3 \cH \phi' + \phi'' \right) + \frac{1}{2} g_{,\phi\phi} \frac{ \phi ^{\prime 4}}{a ^6} \right) \Phib \;.\label{diag}\\
\mathbf{0} & \mathbf{=} & \mathbf{a ^{2} \Big( \delta G \indices{ ^{i} _{j}} - \delta T \indices{ ^{i} _{j}}\Big) ^{k}}  : \nonumber\\
0 & = & - \mathbf{A} - \frac{1}{a ^2} \mathbf{B}' - 2 \frac{\cH}{a ^2} \mathbf{B} + \psib + \frac{1}{a ^{4}} \mathbf{E}'' \;.\label{offdiag}
\end{eqnarray}
These equations may be simplified. First, \eqref{offdiag} reduces \eqref{diag} to
\begin{align}
0 = & \left( 2 \cH + g( \phi) \frac{ \phi ^{\prime 3}}{a ^6} \right) \mathbf{A}' + \Big[ 4 \cH' - 6 \cH ^2 + P_{,X} \phi ^{\prime 2} + \frac{ \phi ^{\prime 2}}{a ^6} \left(g( \phi) (-12 H \phi' + 4 \phi'') + 2 g_{,\phi}\phi ^{\prime 2} \right) \Big] \mathbf{A} \nonumber\\
& + 2 \psib'' - g( \phi) \frac{ \phi ^{\prime 2}}{a ^6} \Phib''  - \left(P_{,X} \phi' + g( \phi)\frac{ \phi'}{a ^{6}} \left(-9H \phi' + 2 \phi'' \right) + 2 g_{,\phi} \frac{ \phi ^{\prime 3}}{a ^{6}} \right) \Phib' \nonumber\\
& - \left(a ^6 P_{, \phi} + g_{,\phi} \frac{ \phi ^{\prime 2}}{a ^{6}} \left(-3 \cH \phi' + \phi'' \right) + \frac{1}{2} g_{,\phi\phi} \frac{ \phi ^{\prime 4}}{a ^6} \right) \Phib \;.\label{diagreduced1}
\end{align}
Next we combine \eqref{bgharm1} and \eqref{bgharm2} to get
\be\label{combfriedmannharmonic}
-\cH'+3\cH^2=\frac12P_{,X}\phi'^2-3\frac{g\cH\phi'^3}{a^6}+\frac12\frac{g_{,\phi}\phi'^4}{a^6}+\frac12\frac{g\phi'^2\phi''}{a^6} \ .
\ee
We can insert \eqref{combfriedmannharmonic} into \eqref{diagreduced1} and obtain
\begin{align}
0 = & \left( 2 \cH + g( \phi) \frac{ \phi ^{\prime 3}}{a ^6} \right) \mathbf{A}' + \Big[ 2 \cH' + \frac{ \phi ^{\prime 2}}{a ^6} \left(g( \phi) (-6 H \phi' + 3 \phi'') +  g_{,\phi} \phi ^{\prime 2} \right) \Big] \mathbf{A} \nonumber\\
& + 2 \psib'' - g( \phi) \frac{ \phi ^{\prime 2}}{a ^6} \Phib''  - \left(P_{,X} \phi' + g( \phi)\frac{ \phi'}{a ^{6}} \left(-9H \phi' + 2 \phi'' \right) + 2 g_{,\phi}\frac{ \phi ^{\prime 3}}{a ^{6}} \right) \Phib' \nonumber\\
& - \left(a ^6 P_{, \phi} + g_{,\phi} \frac{ \phi ^{\prime 2}}{a ^{6}} \left(-3 \cH \phi' + \phi'' \right) + \frac{1}{2} g_{,\phi\phi} \frac{ \phi ^{\prime 4}}{a ^6} \right) \Phib \;.\label{diagreduced2}
\end{align}
If we now take the derivative of \eqref{t-i} and replace in this derivative expression the background equation \eqref{bgharm3}, we directly obtain \eqref{diagreduced2}, hence this equation is redundant.
In \eqref{t-t}, we can also use \eqref{combfriedmannharmonic} to eliminate the $\cH^2$ term. The gauge conditions \eqref{eq:finalPert1} and \eqref{eq:finalPert2} become dynamical equations for $A$ and $B$. Eqn \eqref{eq:finalPert1} can be used to reduce \eqref{offdiag}. Then we are left with the set of equations \eqref{eq:finalPert1} - \eqref{eq:finalPert5} that are used in the main part of the text. These equations are sufficient to determine the evolution of perturbations in the harmonic gauge.

For completeness, we note that the linearized scalar field equation (which is not independent, as it can be derived from the linearized Einstein equations) can be simplified by using \eqref{eq:finalPert5} to replace the $E''$ term, \eqref{eq:finalPert1} to replace the $g\cH\phi'^2\psi'$ term and then \eqref{eq:finalPert2}. Then we obtain
\begin{align}
0  = & \left(P_{,XX} \phi ^{\prime 3} - 12 g( \phi) \cH \phi ^{\prime 2} + 2 g_{,\phi} \phi ^{\prime 3} \right) \mathbf{A}' + \Big[ 2 a ^6 P_{,X} \phi'' + P_{, X \phi} a ^6 \phi ^{\prime 2} + P_{,XX} \phi ^{\prime 2} \left(5 \phi'' - 12 \cH \phi' \right) \nonumber\\
&+ P_{, XX \phi} \phi ^{\prime 4} + P_{, XXX} \frac{ \phi ^{\prime 4}}{a ^{6}} \left( \phi'' - 3 \cH \phi' \right) + g( \phi) \phi' (- 24 \cH \phi'' + 72 \cH ^2 \phi' - 12 \cH' \phi' ) \nonumber\\
& + 8 g_{,\phi} \phi ^{\prime 2} \left( \phi'' - 3 \cH \phi' \right)  + 2 g_{,\phi\phi} \phi ^{\prime 4} \Big] \mathbf{A} +gk^2a^4\phi'^2\psib+ 2g( \phi) k^2a ^2 \phi'\phi''\mathbf{B} - 3 g( \phi) \phi ^{\prime 2} \psib'' \nonumber\\
&- 6 g( \phi) \phi' \phi'' \psib' -2 g( \phi) k^2\phi' \phi''\mathbf{ E}' + \left(- a ^6 P_{,X} - P_{,XX} \phi ^{\prime 2} + 6 g( \phi) \cH \phi' - 2 g_{,\phi}\phi ^{\prime 2} \right) \Phib'' \nonumber\\
&  + \Big[ - a ^6 P_{,X \phi} \phi' - P_{,XX} \phi' \left(3 \phi'' - 9 \cH \phi' \right)  - P_{,XX \phi} \phi ^{\prime 3} - P_{,XXX} \frac{ \phi ^{\prime 3}}{a ^6} \left( \phi'' - 3 \cH \phi' \right) \nonumber\\
&  +2 g( \phi) \left(3\cH \phi'' - 18 \cH ^2 \phi' + 3 \phi' \cH'  \right) - g_{,\phi} \phi' ( 4 \phi'' - 18 \cH \phi') - 2 g_{,\phi\phi} \phi ^{\prime 3}\Big] \Phib' \nonumber\\
&+ \Big[ -a ^{10} k ^2 P_{,X} + a ^{12} P_{, \phi \phi} - a ^{6} P_{, X \phi} \phi'' - a ^{6} P_{, X \phi \phi } \phi ^{\prime 2} - P_{, X X \phi} \phi ^{\prime 2} ( \phi'' - 3 \cH \phi' ) \nonumber\\
&  + 2 k ^2 a ^{4} g( \phi) ( \phi'' - \cH \phi') - g_{,\phi} \phi' ( - 6 \cH \phi'' + 18 \cH ^2 \phi' - 3 \cH ' \phi' ) \nonumber\\
&  - 2 g_{,\phi\phi} \phi ^{\prime 2} ( \phi'' - 3 \cH \phi') - \frac{1}{2} g_{,\phi\phi\phi} \phi ^{\prime 4} \Big] \Phib \;.\label{scalarreduced2}
\end{align}


\bibliographystyle{apsrev}
\bibliography{BIBperturbationsBounce}

\end{document}